\pdfoutput=1
\documentclass[12pt]{article}

\usepackage{amsmath,amssymb,graphicx} 
\usepackage{epsf}
\usepackage{pstricks}
\usepackage{cite}
\usepackage{multirow}
\usepackage{youngtab}

\def\matrixone{\hbox{$1\hskip -1.2pt\vrule depth 0pt height 1.6ex width 0.7pt
\vrule depth 0pt height 0.3pt width 0.12em$}}

\newcommand{\beq}{\begin{eqnarray}}
\newcommand{\eeq}{\end{eqnarray}}

\newcommand{\centeron}[2]{{\setbox0=\hbox{#1}\setbox1=\hbox{#2}\ifdim
                                        
\wd1>\wd0\kern.5\wd1\kern-.5\wd0\fi
\copy0

\kern-.5\wd0\kern-.5\wd1\copy1\ifdim\wd0>\wd1
                                       \kern.5\wd0\kern-.5\wd1\fi}}
\newcommand{\ltap}{\>\centeron{\raise.35ex\hbox{$<$}}
                               {\lower.65ex\hbox{$\sim$}}\>}
\newcommand{\gtap}{\>\centeron{\raise.35ex\hbox{$>$}}
                               {\lower.65ex\hbox{$\sim$}}\>}

\newcommand\ZZ{\hbox{\zfont Z\kern-.4emZ}}
\font\zfont = cmss10 

\begin{document}
\begin{titlepage}
\begin{flushright}
\end{flushright}

\vskip.5cm
\begin{center}
{\huge \bf 
Fundamental Composite 2HDM:\\
SU(N) with 4 flavours
}

\vskip.1cm
\end{center}
\vskip0.2cm

\begin{center}
{\bf
{Teng Ma}$^{a}$, 
 {\rm and}
{Giacomo Cacciapaglia}$^{b}$}
\end{center}
\vskip 8pt

\begin{center}
$^{a}$ {\it Department of Physics, Tsinghua University, Beijing, 100084 China;\\
Center for High energy Physics, Tsinghua University, Beijing, 100084, China} \\
\vspace*{0.1cm}
$^{b}$ {\it Universit\'e de Lyon, F-69622 Lyon, France; Universit\'e Lyon 1, Villeurbanne;\\
CNRS/IN2P3, UMR5822, Institut de Physique Nucl\'eaire de Lyon\\
F-69622 Villeurbanne Cedex, France } \\
\vspace*{0.1cm}
\end{center}

\vglue 0.3truecm

\begin{abstract}
\vskip 3pt
\noindent
{ We present a new model of composite Higgs based on a gauged SU(N) group with 4 Dirac fermions in the fundamental representation. At low energy, the model has a global symmetry SU(4)$\times$SU(4) broken to the diagonal SU(4), containing 2 Higgs doublets in the coset. We study in detail the generation of the top mass via 4-fermion interactions, and the issue of the vacuum alignment. In particular, we prove that, without loss of generality, the vacuum can always be aligned with one doublet. Under certain conditions on the top pre-Yukawas, the second doublet, together with the additional triplets, is stable and can thus play the role of Dark Matter. This model can therefore be an example of composite inert-2HDM model.}

\end{abstract}

\end{titlepage}

\newpage


\section{Introduction}
\label{sec:intro}
\setcounter{equation}{0}
\setcounter{footnote}{0}

A consistent mechanism to provide mass to gauge bosons was proposed in 1964 by Brout, Englert and Higgs~\cite{Englert:1964et,Higgs:1964ia,Higgs:1964pj}, based on the concept of spontaneous symmetry breaking. Once the mechanism is realised in terms of scalar fields, besides the massless Goldstone boson eaten by the massive gauge bosons, the spectrum typically contains massive degrees of freedom~\cite{Higgs:1964pj}: in the case of the Standard Model (SM), this sector consists of a single neutral state, aka the Higgs boson.
Its discovery in 2012 by ATLAS~\cite{Aad:2012tfa} and CMS~\cite{Chatrchyan:2012xdj} at the Large Hadron Collider (LHC), coming 50 years after its theoretical proposal, can be considered the crowning of a long-standing physics program.

The outstanding experimental results have obtained a precise determination of the mass of the new resonance~\cite{Aad:2015zhl}, however the measurement of its couplings, which is the ultimate test of the SM predictions, has been achieved with limited accuracy~\cite{Khachatryan:2014jba,ATLAS:2015bea}. While the central values seem to suggest that the SM hypothesis is correct, the precision attained is only at the level of 10\% in the best channels ($WW$ and $ZZ$). This precision is a far cry from the one attained in other observables of the electroweak sector, where precisions at the level of 0.1\% are common~\cite{ALEPH:2005ab,Schael:2013ita}. From the experimental data, therefore, there is still ample space for extensions of the Higgs sector of the theory, and one may still expect new particles to be present at mass scales not far from the TeV scale. This expectation is also corroborated by theoretical considerations, mainly based on the stability of the Higgs mass, and of the electroweak scale, under quantum corrections -- the infamous hierarchy problem. Furthermore, the SM fails to provide a candidate of Dark Matter, and to explain Baryogenesis (the generation of matter-antimatter asymmetry in the Universe).

Extensions of the Higgs sector of the SM are often required in models of New Physics addressing the hierarchy problem. One very attractive possibility is to replace the elementary scalar at the origin of the symmetry breaking in the SM with a confining sector which spontaneously breaks the symmetry via confinement. Such a physical effect does occur in nature in Quantum Chromodynamics (QCD), the confining sector describing nuclear strong interactions. Early attempts were made in the late 70's by scaling up QCD dynamics~\cite{Dimopoulos:1979es} to energies apt to generate the electroweak scale (old school Technicolor), however such theories did not have a light Higgs boson and typically induced too large corrections to electroweak precision measurements~\cite{Peskin:1990zt}. One way to introduce a Higgs-like boson is to extend the global symmetry of the model so that a light scalar can be left in the spectrum as a pseudo-Nambu-Goldstone boson (pNGB)~\cite{Kaplan:1983fs,Kaplan:1983sm}.
This idea received new life in the early 90's when, following the conjecture of a duality between warped extra dimensions and 4-dimensional conformal theories, a pNGB Higgs was associated with a gauge field in extra dimensions (holographic Higgs)~\cite{Contino:2003ve}. The first concrete and feasible model was proposed in~\cite{Agashe:2004rs,Agashe:2005dk}, based on the minimal coset SO(5)/SO(4) that provides just a SM-like Higgs with custodial symmetry. A lot of work has been dedicated to this class of models (for recent reviews, see~\cite{Bellazzini:2014yua} and~\cite{Panico:2015jxa}), however most of the work has been dedicated to the minimal scenario and formulated in an effective field theory context. Furthermore, following the holographic Higgs construction, the model building efforts have been relying on the presence of fermionic bound states (top partners) which couple linearly to the SM fermions (the top) in an attempt to give the top mass without incurring in large flavour violating effects.

In this paper we will take a different approach to the problem: instead of relying on an effective Lagrangian (possibly completed by a conformal theory) or extra dimensions, we will define an Ultra--Violet (UV) completion based on a simple confining gauge group with fundamental fermions. Relying on a Fundamental Composite Dynamics (FCD) allows us to draw a precise relation between the components of the underlying model and the composite states present in the spectrum of the effective theory. Furthermore, one can study the relation between the limit in which the Higgs appears as a pNGB, and a Technicolor-like limit of the theory~\cite{Cacciapaglia:2014uja}. The minimal FCD model, based on a gauged SU(2)$_{\rm FCD}$ confining dynamics with 4 Weyl fermions in the fundamental representation~\cite{Ryttov:2008xe,Galloway:2010bp}, enjoys a global symmetry SU(4) broken to Sp(4). The symmetry breaking pattern has been confirmed on the Lattice~\cite{Hietanen:2014xca}. The phenomenology of the scalar sector, which contains an extra singlet, has been recently studied in~\cite{Arbey:2015exa}.
This example shows that an extended Higgs sector is typically predicted in composite models with an underlying FCD. We are thus interested in exploring less minimal possibilities, with a two-fold purpose. On one hand, the pNGB scalars are the lightest particles one would naturally expect in this class of models, thus it is of paramount importance to establish the capability of the LHC to discover them or probe their existence. On the other hand, larger symmetry groups can enjoy unbroken discrete symmetries that may protect some of the pNGBs, thus providing a natural composite Dark Matter candidate. This possibility has been studied in the literature in the SU(4)/Sp(4) case in the effective field theory context~\cite{Frigerio:2012uc,Marzocca:2014msa}, however only having a UV completion allows us to determine the stability of the pNGB. In fact, Wess-Zumino-Witten anomaly terms~\cite{Wess:1971yu,Witten:1983tw}, generated by the fermionic components of the composite scalar, may induce prompt decays into a pair of gauge bosons: this indeed occurs in the minimal case~\cite{Galloway:2010bp,Cacciapaglia:2014uja,Gripaios:2009pe}.

 We focus here on the case of a global symmetry SU(4)$\times$SU(4) broken to the diagonal SU(4), which can be obtained from a FCD based on the confining gauge group SU($N$)$_{\rm FCD}$ with 4 Dirac fermions in the fundamental representation.
The fermion multiplicity 4 is the minimal one required in order to have a Higgs-like state among the pNGBs and custodial symmetry~\footnote{Note that SU(3)$\times$SU(3) would also contain a Higgs-like state, but without custodial symmetry.}, as SO(4)$\subset$SU(4).
This same symmetry breaking pattern has been used in the construction of a Little Higgs (as the isomorphic SO(6)$\times$SO(6)) in~\cite{Schmaltz:2010ac}.
A nice feature of this model is that it contains two electroweak doublets in the pNGB spectrum, thus giving rise to a 2 Higgs Doublet Model (2HDM): a first general analysis of composite 2HDMs can be found in~\cite{Mrazek:2011iu}, where the authors focus on symmetries with the minimal cosets.
Contrary to supersymmetry or standard 2HDMs, where both doublets can acquire a vacuum expectation value independently on their couplings, in composite scenarios the structure of the vacuum depends on the couplings to the fermions (in particular, the top).
This fact derives from the loop induced potential for the Higgs, which is generated by explicit breaking terms of the global symmetry, like the Yukawa couplings. We will study in detail the vacuum alignment mechanism, using the UV completion as a guiding line.
In our model, the fermion masses come directly from four-fermion interactions bilinear in the elementary fields: we will pragmatically assume that the physics responsible for generating such interactions does not induce too large flavour changing neutral currents. This is a non trivial assumption~\cite{Eichten:1979ah}, however the eventual solution to the flavour puzzle should not affect the Higgs potential and low energy pNGB Lagrangian. 
In modern incarnations of composite pNGB Higgs models, the flavour puzzle is partially addressed by the mechanism of partial compositeness~\cite{Kaplan:1991dc}, inspired by the flavour protection in models on warped extra dimensions~\cite{Gherghetta:2000qt,Huber:2000ie}. Partial compositeness, however, requires the presence of light coloured fermionic bound states in the low energy spectrum, and obtaining such states in a FCD model can be quite challenging~\cite{Ferretti:2013kya}. Furthermore, in the 4-dimensional model, an explanation of the origin of the mixing terms, which are typically related to four-fermion interactions, is missing.

One of the main advantages of  FCD formulations of composite Higgs model is that it allows for Lattice simulation to study the spectrum and behaviour of the dynamical model: the case of SU(3)$_{FCD}$ with 4 flavour in the fundamental representation has been studied and shown to condense~\cite{Hasenfratz:2009ea,Fodor:2009wk,Aoki:2012eq,Aoki:2013xza}, as expected. Furthermore, perturbative arguments indicate that the model is outside of the conformal window, thus expected to condense, for any value of the FCD colours $N$~\cite{Dietrich:2006cm}.

The paper is organised as follows: in Section~\ref{sec:model1}, after presenting the general set up of the model, we discuss in detail the alignment of the vacuum and constraints from the Higgs couplings and electroweak precision measurements. Then in Section~\ref{sec:DM} we discuss the conditions allowing for the presence of a Dark Matter candidate. In Section~\ref{sec:lattice} we discuss the spectrum of heavier states, before the concluding remarks.

\section{The model: SU($N$) with 4 Dirac flavours in the fundamental}
\label{sec:model1}
\setcounter{equation}{0}
\setcounter{footnote}{0}

\begin{table}[tb]
\begin{center}
\begin{tabular}{c|ccc|}
    & SU($N$) & SU(2)$_L$ & U(1)$_Y$ \\
\hline
\multirow{2}{*}{$ \psi_L = \left( \begin{array}{c}  \psi_1  \\  \psi_2 \end{array} \right)$}& \multirow{2}{*}{${\tiny{\yng(1)}}$} & \multirow{2}{*}{\bf 2}  & \multirow{2}{*}{0} \\
 & &  & \\
\multirow{2}{*}{$ \psi_R = \left( \begin{array}{c}  \psi_3  \\  \psi_4 \end{array} \right)$} & \multirow{2}{*}{${\tiny{\yng(1)}}$}  & {\bf 1} & 1/2 \\
 & & {\bf 1} & -1/2  \\
\hline
\end{tabular} 
\caption{Quantum numbers of the fundamental fermion under the confining FCD group SU($N$), and the electroweak group SU(2)$_L \times$ U(1)$_Y$. The values refer to the left-handed component of the Dirac fermion.} \label{tab:QN}
\end{center} \end{table}

The model is based on a strongly interacting SU($N$)$_{\rm FCD}$ group with 4 Dirac fermions $\psi_i$ in the fundamental representation.
The electroweak (EW) symmetry is embedded by assigning electroweak quantum numbers to the fundamental fermions (techni-fermions), as detailed in Table~\ref{tab:QN}. The custodial symmetry SO(4) $\sim$ SU(2)$_L \times$ SU(2)$_R$ is realised by $\psi_{3,4}$ forming a doublet of SU(2)$_R$ (with the hypercharge associated to the diagonal generator of SU(2)$_R$, as usual). 
As the couplings are vector-like, no gauge anomalies are introduced, so that the model is consistent.
The global symmetry of the strong sector is SU(4)$_1\times$ SU(4)$_2 \times$ U(1)$_{TB}$.
The non-anomalous U(1)$_{TB}$ corresponds to the techni-baryon (TB) number, which is conserved in this model.
The Lagrangian, to be added to the SM one, is
\beq
\mathcal{L}_{FCD} = i \bar{\psi} D_\mu \gamma^\mu \psi - \bar{\psi} M_Q \psi\,.
\eeq
The mass $M_Q$ is invariant under the SM custodial symmetry:
\beq
M_Q = \left( \begin{array}{cc}
m_L & 0 \\
0 & m_R
\end{array} \right)\,,
\eeq
i.e., the masses of $\psi_3$ and $\psi_4$ are chosen to be equal to preserve SU(2)$_R$.
The covariant derivative contains both the FCD gauge interactions, and the EW gauge interactions, which are embedded in the diagonal SU(4)$_D$ as:
\beq
T_L^i = \frac{1}{2} \left( \begin{array}{cc}
\sigma_i & 0 \\
0 & 0
\end{array} \right)\,, \quad T_R^i = \frac{1}{2} \left( \begin{array}{cc}
0 & 0 \\
0 & \sigma_i 
\end{array} \right)\,.
\eeq
Note that the hypercharge is given by $Y = T_R^3$.

The FCD dynamics leads to condensation in the infrared: it is convenient to analyse the model by using Weyl fermions
\beq
\psi = \left( \begin{array}{c}
\chi \\
\bar{\eta}
\end{array} \right)
\eeq
where $\chi$ (the left-handed chirality) transforms as a ${\bf 4}$ of SU(4)$_1$ and $\eta$ (the right-handed chirality) as a ${\bf \bar{4}}$ of SU(4)$_2$.
The condensate transforms as
\beq
\langle \eta \chi \rangle = ({\bf 4}, {\bf \bar{4}})_{\rm SU(4)_1 \times SU(4)_2}
\eeq
and it breaks SU(4)$_1\times$ SU(4)$_2 \to$ SU(4)$_D$. The condensate has no TB charge.
This breaking entails 15 Goldstone Bosons, transforming as the adjoint of the unbroken SU(4)$_D$.
We can first align the vacuum along the direction that does not break the EW symmetry, i.e.
\beq
\langle \eta \chi \rangle = \left( \begin{array}{cc}
\matrixone & 0 \\
0 & \matrixone
\end{array} \right)\,,
\eeq
which is aligned with the SU(4)$_D$ preserving techni-fermion mass ($m_L = m_R$).
The 15 pNGBs transform under the custodial symmetry SU(2)$_L \times$ SU(2)$_R$ as
\beq
{\bf 15}_{\rm SU(4)_D} = (2,2) + (2,2) + (3,1) + (1,3) + (1,1)\,;
\eeq
the model therefore contains two doublets that may play the role of the Brout-Englert-Higgs doublet field.
The pion matrix can now be parametrised as:
\beq
\Pi =  \frac{1}{2} \left( \begin{array}{cc}
\sigma_i \Delta^i + s/\sqrt{2}  &  -i \Phi_H \\
i \Phi_H^\dagger & \sigma_i N^i - s/\sqrt{2}
\end{array} \right)
\eeq
with
\beq
\sigma_i \Delta^i = \left( \begin{array}{cc}
\Delta_0  & \sqrt{2} \Delta^+ \\
\sqrt{2} \Delta^- & - \Delta_0
\end{array} \right)\,, \qquad \sigma_i N^i = \left( \begin{array}{cc}
 N_0  & \sqrt{2} N^+ \\
\sqrt{2} N^- & - N_0
\end{array} \right)\,,
\eeq
being the two triplets of SU(2)$_L$ and SU(2)$_R$ respectively, $s$ the singlet, and $\Phi_H$ containing the two bi-doublets $H_{1,2}$:
\beq
\Phi_H = \left( i \sigma_2 (H_1^\ast + i H_2^\ast), \;\; H_1 + i H_2 \right)\,.
\eeq
We can already note a special feature of this model: the two Higgs doublets appear as a complex bi-doublet of the custodial symmetry, and this fact will have important consequences for the vacuum structure.
The pion matrix is then embedded in 
\beq
U = e^{i \Pi/f}
\eeq
transforming linearly under the stability group SU(4)$_D$ as $U \to \Omega_D \cdot U \cdot \Omega_D^\dagger$\,.

The vacuum where the EW symmetry is broken can be though of as being generated by a vacuum expectation value (VEV) for the two doublets:
\beq
\langle \Phi_H \rangle = \frac{v}{\sqrt{2}} e^{i \beta}\; \matrixone \label{eq:vevI}
\eeq
where $v = \sqrt{v_1^2 + v_2^2}$ and $\tan \beta = \frac{v_2}{v_1}$, $v_{1,2}$ being the VEVs of the two doublets. Note that the VEV in Eq.(\ref{eq:vevI}) is the most general one that preserves the custodial symmetry: any other choice would contribute to the $\rho$ parameter at tree level. This effect is more properly described as a misalignment of the vacuum generated by a symmetry of the broken generators. In this case, we have:
\beq \label{eq:Omega}
\Omega (\theta, \beta) = \left( \begin{array}{cc}
\cos \frac{\theta}{2}  & e^{i \beta} \sin \frac{\theta}{2}  \\
- e^{-i \beta} \sin \frac{\theta}{2}  & \cos \frac{\theta}{2} 
\end{array} \right)\,, \quad \Sigma_1 = \Omega \cdot \Omega = \Omega (2\theta, \beta)\,,
\eeq
with $v = 2 \sqrt{2} f \theta$, and $\Sigma_1$ the properly aligned vacuum.
As the symmetry breaking pattern is unaltered, the pion matrix contains the same number of Goldstone bosons, which, in the new vacuum, can be parametrised as the linearly transforming matrix
\beq \label{eq:Sigma}
\Sigma = \Omega (\theta, \beta) \cdot U \cdot \Omega (\theta, \beta)\,.
\eeq

At leading order, the chiral Lagrangian is given by the kinetic term for $\Sigma$:
\beq
\mathcal{L}_{CCWZ} = f^4 \mbox{Tr} [(D_\mu \Sigma)^\dagger D^\mu \Sigma]\,. \label{eq:CCWZI}
\eeq
This term contains mass terms for the $W$ and $Z$
\beq
m_W^2 = 2 g^2 f^2 \sin^2 \theta\,, \quad m_Z^2 = \frac{m_W^2}{\cos^2 \theta_W}\,.
\eeq
so that
\beq
2 \sqrt{2} f \sin \theta = v_{\rm SM} = 246\; \mbox{GeV}\,.
\eeq
The Goldstone Bosons eaten by the massive $W$ and $Z$ are
\beq
\pi^\pm = \cos \beta\ H_1^\pm + \sin \beta\ H_2^\pm\,, \quad \pi_0 = \sqrt{2}\, \mbox{Im} [\cos \beta\ H_1^0 + \sin \beta\ H_2^0]\,,
\eeq
and, following the usual notation in 2HDMs, we define the physical scalars as:
\beq
H^\pm = - \sin \beta\ H_1^\pm + \cos \beta\ H_2^\pm\,, \quad A_0 = \sqrt{2}\, \mbox{Im} [-\sin \beta\ H_1^0 + \cos \beta\ H_2^0]\,, \nonumber \\
 h_1 = \sqrt{2}\, \mbox{Re} [\cos \beta\ H_1^0 + \sin \beta\ H_2^0]\,, \quad h_2 = \sqrt{2}\, \mbox{Re} [-\sin \beta\ H_1^0 + \cos \beta\ H_2^0]\,. \label{eq:basis1}
\eeq
The only field with linear couplings to the gauge bosons is $h_1$, which thus can play the role of the Higgs boson.
Its couplings are given by
\beq
g_{WWh_1} = \cos^2 \theta_W g_{ZZh_1} = \sqrt{2} g^2 f \sin \theta \cos \theta = \frac{2 m_W^2}{v_{\rm SM}} \cos \theta = g_{WWh}^{\rm SM} \cos \theta\,.
\eeq
The couplings of two scalars to gauge bosons are reported in the Appendix~\ref{app:couplings}: we note here that none of the couplings depend on $\beta$.
In fact, the parameter $\beta$ can be rotated away from the vacuum by using the transformation
\beq \label{eq:Omegabeta}
\Omega_\beta = \mbox{Exp} \left[ - i \frac{\beta}{2} \left( \begin{array}{cc}
1 & 0 \\ 0 & -1 \end{array} \right) \right] = \left( \begin{array}{cc}
e^{-i \beta/2} & 0 \\ 0 & e^{i \beta/2} \end{array} \right)
\eeq
so that
\beq
\Sigma_1 (\beta=0) = \Omega_\beta \cdot \Sigma_1 \cdot \Omega_\beta^\dagger\,.
\eeq
As the gauge interactions (and the techni-fermion mass) are left invariant under this transformation, the Lagrangian in Eq.(\ref{eq:CCWZI}) is independent on $\beta$, once the pion fields are properly re-labeled as in Eq.~\ref{eq:basis1}.
The transformation in Eq.~\ref{eq:Omegabeta} is generated by a U(1) symmetry which is unbroken in the EW-preserving vacuum: under such symmetry, the complex bi-doublet  $\Phi_H$ is charged, while the triplets and singlet are neutral. 

The transformation of the pion matrix under CP can be obtained by the composition of the scalars in terms of fundamental fermions:
\beq
CP(\Sigma) = \Sigma^\ast (-\overrightarrow{x})\,; \quad CP(A_\mu) = (-1)^{\delta^{\mu 0}} A_\mu\,, \quad CP(x^\mu) = - (-1)^{\delta^{\mu 0}} x^\mu\,,
\eeq
where the gauge vector, and space-time co-ordinates, are CP-transformed in the standard way. 
From the above definition, it is clear that $\beta$ is a CP-odd parameter, i.e. it violates CP invariance. Thus, one can define the CP properties of the pNGBs only in the case $\beta = 0$.
As usual when writing an effective Lagrangian for Goldstone bosons, it is possible to define an intrinsic parity of the pion, dubbed Goldstone parity (GP), which acts on the pion matrix as:
\beq
GP(\Sigma) = P_{\rm GP} \cdot \Sigma^\dagger (-\overrightarrow{x}) \cdot P_{\rm GP}^\dagger\,,
\eeq
while the gauge vectors and co-ordinates are CP-transformed, and
\beq
P_{\rm GP} = \left( \begin{array}{cc}
\sigma_2 & 0 \\
0 & -\sigma_2
\end{array} \right)\,.
\eeq
As $P_{\rm GP} \cdot \Omega^\dagger (\theta, \beta) \cdot P_{\rm GP}^\dagger = \Omega (\theta, \beta)$, it is clear that this time $\beta$ is a GP-even parameter.
The transformation under GP, and CP for $\beta = 0$, of the pNGBs are summarised in Table~\ref{tab:GPCP}: we see that under CP, it is the singlet $s$, the triplets, and $h_2$ that transform as pseudo-scalar fields. On the other hand, under GP, which is compatible with a non-zero value of $\beta$, it is $s$ and $A_0$ to be odd, like in more traditional 2HDM models.

\begin{table}[tb] \begin{center}
\begin{tabular}{c|cccccc|ccc}
  & $h_1$ & $h_2$ & $A_0$ & $s$ & $\Delta_0$ & $N_0$ & $H^\pm$ & $\Delta^\pm$ & $N^\pm$ \\
\hline
CP ($\beta = 0$) & $+$ & $-$ & $+$ & $-$ & $-$  & $-$ & $-$ & $-$ & $-$ \\
\hline
GP & $+$ & $+$ & $-$ & $-$ & $+$  & $+$ & $+$ & $+$ & $+$ 
\end{tabular}
\caption{Parities under CP (for $\beta = 0$) and GP of the pNGBs: for the charged states, it is left understood that they transform in their complex conjugates (anti-particles).} \label{tab:GPCP}
\end{center} \end{table}

\subsection{Vacuum alignment part 1: top Yukawa couplings fix $\beta$} \label{sec:vac1}

The alignment of the condensate in the flavour space is determined by the explicit symmetry breaking terms: in the minimal model, they are the mass of the techni-fermions $M_Q$, the gauge couplings and the terms giving mass to the SM fermions.
The last two generate a potential via loops.
As the mass and gauge interactions are invariant under the symmetry in Eq.~\ref{eq:Omegabeta}, only the top loops may be sensitive to the value of $\beta$.
We will therefore concentrate first on the effect of the top mass on the vacuum alignment, and discuss the alignment in the full potential in the next section.

We will assume here that the top Yukawa couplings are generated via 4-fermion operators connecting the elementary quarks to the techni-fermions, which are bilinear in the elementary fields. The possibility of generating the top mass via partial compositeness will be considered in a future work (see also~\cite{Vecchi:2015fma}). The couplings are generated by an unspecified physics at a scale $\Lambda_t \gtrsim 4 \pi f$, and in most generality 4 terms can be written down:
\beq
\mathcal{L}_{4-Fermi} &=& - \frac{1}{\Lambda_t^2} (\bar{\chi}_{q_L} \bar{\eta}_{t_R})^\alpha \left[ \tilde{y}_{t1} (\eta^{up}_R \chi_L)_\alpha   +  \tilde{y}_{t2} (\bar{\chi}_R^{up} \bar{\eta}_L )_\alpha  +  \right. \nonumber \\
& & \left.  \tilde{y}_{t3}  (i\sigma_2)_{\alpha \beta}(\eta_L \chi_R^{down})^\beta  +  \tilde{y}_{t4}  (i\sigma_2)_{\alpha \beta}(\bar{\chi}_L \bar{\eta}_R^{down})^\beta    \right] + h.c. \label{eq:4Fermi}\\
&=& - \frac{1}{\Lambda_t^2} (\bar{\chi}_{q_L} \bar{\eta}_{t_R})^\alpha  \left[ \left(\tilde{y}_{t1} P_{1,\alpha} + \tilde{y}_{t3} (i\sigma_2)_{\alpha \beta} P_2^\beta \right)^{ij} \eta_j \chi_i + \right. \nonumber \\
& & \left. \left(\tilde{y}_{t2} P_{1,\alpha} + \tilde{y}_{t4} (i\sigma_2)_{\alpha \beta} P_2^\beta \right)^{ij} \bar{\chi}_j \bar{\eta}_i \right] + h.c.
\eeq
where $\chi_{q_L}$ and $\eta_{t_R}$ are Weyl fermions corresponding to the elementary doublet and singlet containing the top, $\alpha$ and $\beta$ are indices in the gauged SU(2)$_L$, and $i$ ($j$) is an index in SU(4)$_1$ (SU(4)$_2$).
In the second line, we have embedded the couplings in the full flavour SU(4)$_1\times$SU(4)$_2$ space by use of spurion matrices $P_1$ and $P_2$, which transform as a doublet and an anti-doublet of SU(2)$_L$, defined as:
\beq
P_{1,1} = \left( \begin{array}{cccc}
0 & 0 & 0 & 0\\
0 & 0 & 0 & 0\\
1 & 0 & 0 & 0\\
0 & 0 & 0 & 0
\end{array} \right)\,, \qquad P_{1,2} = \left( \begin{array}{cccc}
0 & 0 & 0 & 0\\
0 & 0 & 0 & 0\\
0 & 1 & 0 & 0\\
0 & 0 & 0 & 0
\end{array} \right)\,; \nonumber \\
P_2^1 = \left( \begin{array}{cccc}
0 & 0 & 0 & 1\\
0 & 0 & 0 & 0\\
0 & 0 & 0 & 0\\
0 & 0 & 0 & 0
\end{array} \right)\,, \qquad P_2^2 = \left( \begin{array}{cccc}
0 & 0 & 0 & 0\\
0 & 0 & 0 & 1\\
0 & 0 & 0 & 0\\
0 & 0 & 0 & 0
\end{array} \right)\,.
\eeq
The above Lagrangian contains 4 independent complex couplings, the pre-Yukawas $\tilde{y}_{ti}$, however not all the phases are physical. This fact can be easily understood in terms of the 4-fermion interactions in Eq.~\ref{eq:4Fermi}. After fixing the techni-fermion mass terms real, 2 phases can be reabsorbed in an arbitrary phase redefinition of the fermion fields: the relative phase between the SU(2)$_L$ and SU(2)$_R$ doublets and the relative phase between the two SM fields.
The former can be embedded in SU(4) and identified with the following transformations:
\beq
\Omega_\beta = \left( \begin{array}{cc}
e^{-i \beta/2} & 0 \\ 0 & e^{i \beta/2} \end{array} \right)\,,
\eeq
which coincides with the SU(4) transformation in Eq.~\ref{eq:Omegabeta} that allows to remove $\beta$ from the vacuum structure, thus suggesting that $\beta$ may be an unphysical parameter unless the loop-induced potential generates spontaneously a non-vanishing value at the minimum.
The phase redefinition of the SM fields is the usual one that allows to write a real mass for the top.
The phase structure of the pre-Yukawa couplings is crucial as it determines the alignment of the vacuum: we will therefore use the 2 arbitrary phases to align the vacuum to its simplest form.
 Operatively, minimising the potential allows to determine $\beta$ as a function of the phases in the pre-Yukawas; then we can fix the pre-Yukawa phases, or equivalently apply the phase transformation in Eq.~\ref{eq:Omegabeta}, to set $\beta = 0$ in the vacuum, without loss of generality. This means that vacua with non-vanishing $\beta$ are physically equivalent to the vacuum with $\beta = 0$.

In the effective Lagrangian the Yukawa couplings can be written in the form:
\beq \label{eq:Yuk0}
\mathcal{L}_{\rm Yuk} = - f\, (\bar{\chi}_{q_L} \bar{\eta}_{t_R})^\alpha \left[ \mbox{Tr} [P_{1,\alpha} (y_{t1} \Sigma + y_{t2} \Sigma^\dagger)] + (i\sigma_2)_{\alpha \beta} \mbox{Tr} [P_2^\beta (y_{t3} \Sigma + y_{t4} \Sigma^\dagger)] \right] + h.c.
\eeq
where $y_{ti}$ are related to the 4-fermion couplings $\tilde{y}_{ti}$ via form factors of the dynamics.
Once expanding $\Sigma$, this term will generate a mass for the top, and couplings of the pNGBs to the top and bottom quarks.
To study the effect on the vacuum, we will assume that it acquires the simplest possible form, i.e. Eq.~\ref{eq:Sigma} with $\beta = 0$.
It is convenient to define combinations of the Yukawa couplings as follows:
\beq
Y_t = \frac{y_{t1} - y_{t2} - (y_{t3} - y_{t4})}{2 \sqrt{2}}\,, &  & Y_D = \frac{y_{t1} - y_{t2} + (y_{t3} - y_{t4})}{2 \sqrt{2}}\,, \nonumber \\
Y_T = \frac{y_{t1} + y_{t2} + (y_{t3} + y_{t4})}{2 \sqrt{2}}\,, &  & Y_0 =  \frac{y_{t1} + y_{t2} - (y_{t3} + y_{t4})}{2 \sqrt{2}}\,. 
\eeq
Expanding Eq.~\ref{eq:Yuk0} to linear order in the pNGB fields, we obtain:
\beq \label{eq:Yuk1}
\mathcal{L}_{\rm Yuk} &=& - \left[  Y_t  v_{SM} + Y_t \cos \theta\ h_1 + i Y_D\ h_2 + Y_D \cos \theta\ A_0+ i \frac{Y_T}{\sqrt{2}} \sin \theta\ (N_0 + \Delta_0)\right] (\bar{\chi}_{t_L} \bar{\eta}_{t_R}) \nonumber \\
&& - \left[ -i \sqrt{2} Y_D \cos \theta\ H^- + i Y_T \sin \theta\ (N^- + \Delta^-) \right] (\bar{\chi}_{b_L} \bar{\eta}_{t_R}) + h.c.
\eeq
The above expansion clearly shows how to interpret the various couplings: $Y_t$ corresponds to the effective top Yukawa coupling with $m_t = Y_t v_{SM}$, $Y_D$ is the coupling of the second doublet to the top, $Y_T$ is the coupling of the two triplets~\footnote{The SU(2)$_L$ and SU(2)$_R$ triplets share the same coupling due to custodial symmetry, as they form a 6-plet of SO(4).}.

A loop of tops will generate a potential term for the pNGBs, in the form
\beq \label{eq:Vtop}
V_t = - f^4 C_t \left| \mbox{Tr} [P_{1,\alpha} (y_{t1} \Sigma + y_{t2} \Sigma^\dagger)] + (i\sigma_2)_{\alpha \beta} \mbox{Tr} [P_2^\beta (y_{t3} \Sigma + y_{t4} \Sigma^\dagger)] \right|^2
\eeq
with the appropriate SU(2)$_L$ contractions left understood, and with $C_t$ being a positive coefficient depending on the dynamics.
Expanding around the $\beta = 0$ vacuum up to linear order in the pNGB fields, we obtain:
\begin{multline} \label{eq:Vtop0}
V_t = - f^4 C_t \left(8 |Y_t|^2 \sin^2 \theta + 2 \sqrt{2} |Y_t|^2 \sin (2 \theta) \frac{h_1}{f} - i\, 2 \sqrt{2} (Y_D^\ast Y_t - Y_D Y_t^\ast) \sin \theta \frac{h_2}{f} \right.  \\
 \left. + \sqrt{2} (Y_D^\ast Y_t + Y_D Y_t^\ast) \sin (2\theta) \frac{A_0}{f} - i\, 2 (Y_T^\ast Y_t - Y_T Y_t^\ast) \sin^2 \theta \frac{N_0 + \Delta_0}{f} + \dots \right)\,.
\end{multline}
The presence of tadpoles for the neutral pNGBs implies that the chosen vacuum is not consistent: the tadpole for $h_1$ will be fixed once the proper minimum value for $\theta$ is chosen, in fact the top contribution alone generates a minimum for $\theta = \pi/2$ (corresponding to the Technicolor limit) for which the tadpole vanishes.
For the other pNGBs, the tadpoles need to vanish as they are only generated by top loops.
The tadpole of $h_2$ is correlated to the value of $\beta$ on the vacuum: it is proportional to the $\mbox{Im} (Y_t Y_D^\ast)$, and it can be shown that the phases of $Y_t$ and $Y_D$ are directly related to the 2 arbitrary phases of the 4 Yukawa couplings. In other words, one can always choose $Y_t$ and $Y_D$ to be real by properly fixing the phase of the fundamental techni-fermion fields. Then, the minimisation condition of the potential will fix $\beta$ in the vacuum and the vanishing of the tadpole proves that $\beta = 0$ is the correct value at the minimum.
A general analysis of this condition can be found in Appendix~\ref{app:phases}.
This analysis finally proves that $\beta$ in the vacuum is an unphysical parameter, and in the following we will always work in the vacuum with $\beta = 0$.

The tadpoles of $A_0$ and of the triplets are physically relevant as their presence would force the vacuum to a direction that breaks custodial symmetry. Experimental constraints, especially from electroweak precision measurements, would require their values to be small. In the following we will limit ourselves to a vacuum that is exactly custodial invariant, thus eliminating the two tadpoles imposes non-trivial conditions on the 4 Yukawa couplings.
Using the arbitrary overall phase to render $Y_t$ real and positive, the vanishing of the tadpoles can be obtained by imposing
\beq \label{eq:zerotadpoles}
\mbox{Re} (Y_D) = 0\,, \qquad \mbox{Im} (Y_T) = 0\,,
\eeq
where, as we already discussed, $\mbox{Im} (Y_D)$ can be set to zero without loss of generality.

It is instructive to analyse the two conditions in two simple scenarios: one where the 4-fermion interactions are generated by a spin-1 mediator, {\it \'a la} Extended Technicolor (ETC)~\cite{Eichten:1979ah}, and one where the mediator is a heavy scalar with custodial invariant couplings, {\it \'a la} Bosonic Technicolor (BTC)\cite{Samuel:1990dq,Kagan:1991gh}.

\subsubsection*{``ETC'' Yukawas: spin-1 mediator}

As a spin-1 gauge boson only couples to vector currents, the only pre-Yukawas that can be generated after Fierzing are $\tilde{y}_{t1}$ and $\tilde{y}_{t3}$: in this case, there are only two phases which are both unphysical, so that we can choose all the pre-Yukawas real.
Furthermore
\beq
Y_t = Y_0 = \frac{y_{t1} - y_{t3}}{2 \sqrt{2}}\,, \quad Y_D = Y_T =  \frac{y_{t1} + y_{t3}}{2 \sqrt{2}}\,,
\eeq
and, as the doublet and triplet Yukawas are equal, the vanishing of the tadpole for $A_0$ is enough to ensure the vanishing of the triplet tadpoles.
Custodial invariance in the vacuum can therefore be recovered if the two pre-Yukawas are related:
\beq
y_{t3} = - y_{t1}\,, \quad \Rightarrow \;\; Y_t = Y_0 = \frac{y_{t1}}{\sqrt{2}}\,, \quad Y_D = Y_T = 0\,.
\eeq

\subsubsection*{``BTC'' Yukawas: scalar mediator}

In this case we imagine that the 4-fermion interactions are generated by a heavy scalar field transforming as a real bi-doublet of the custodial SU(2)$_L \times$ SU(2)$_R$, which has Yukawa couplings with both techni-fermions and the elementary quarks.
Only the couplings to the techni-fermions need to be custodial invariant, while the SM quarks couple with the SU(2) doublet component with the correct hypercharge.
This leads to a peculiar structure in the Yukawa couplings:
\beq
y_{t4} = y_{t1}^\ast e^{- 2 i \gamma_0}\,, \quad y_{t3} = y_{t2}^\ast e^{- 2 i \gamma_0}\,,
\eeq
i.e. pair of couplings are one the complex conjugate of the other up to an overall arbitrary phase ($\gamma_0$).
Choosing the overall phase equal to zero, the physical Yukawa couplings are given by
\beq
Y_t = \frac{\mbox{Re} (y_{t1} - y_{t2})}{\sqrt{2}}\,, \quad Y_D = i \frac{\mbox{Im} (y_{t1} - y_{t2})}{\sqrt{2}}\,, \quad Y_T = \frac{\mbox{Re} (y_{t1} + y_{t2})}{\sqrt{2}}\,, \quad Y_0 = i \frac{\mbox{Im} (y_{t1} + y_{t2})}{\sqrt{2}}\,. \nonumber
\eeq
The fact that $Y_D$ is always imaginary, while $Y_t$ and $Y_T$ are real, immediately explains why the tadpole for $A_0$  and the triplets are identically vanishing: this is a direct consequence of the custodial invariance of the couplings of the scalar mediator to the techni-fermions.
Finally, $Y_D = 0$ by choosing the relative phase of the techni-fermions, which thus cancels the relative phase between $y_{t1}$ and $y_{t2}$.
This model can therefore be described in terms of 3 physical parameters: the top Yukawa $Y_t$ (fixed by the top mass value), $Y_T$ and the imaginary parameter $Y_0$.

\subsection{Vacuum alignment part 2: fixing $\theta$ and the pNGB masses}

A potential for the light scalars, which also determines the alignment of the vacuum in the SU(4)$_1 \times$SU(4)$_2$ space, is generated by the explicit breaking of the global symmetry.
In the minimal case, there are only 3 sources of breaking, necessary to have a viable model: the mass of the techni-fermions $M_Q$, the partial gauging of the global symmetry (i.e. the SM gauge couplings), and the 4-fermion interactions generating the top Yukawas.
At leading order, therefore, we can expect 3 main contributions to the potential.

The first comes from the techni-fermion mass terms:
\beq \label{eq:Vmass}
V_{\rm mass} &=& - C_m f^3 \mbox{Tr} [M_Q \cdot \Sigma] + h.c. = \nonumber \\
& & - 4 C_m f^3 (m_L + m_R) \cos \theta + \sqrt{2} C_m f^2 (m_L + m_R) \sin \theta\, h_1 + \dots
\eeq
One loop of EW gauge bosons contributes:
\beq \label{eq:Vgauge}
V_{\rm gauge} &=& - C_g f^4 \left( g^2 \sum_a \mbox{Tr} [T_L^a\cdot \Sigma \cdot T_L^a \cdot \Sigma^\dagger] + {g'}^2 \mbox{Tr} [T_R^3\cdot \Sigma \cdot T_R^3 \cdot \Sigma^\dagger] \right)= \nonumber \\
&& - C_g f^4 \frac{3 g^2 + {g'}^2}{2} \cos \theta^2 + C_g f^3 \frac{3 g^2 + {g'}^2}{4 \sqrt{2}} \sin (2\theta) \, h_1 + \dots
\eeq
The two coefficients $C_m$ and $C_g$ are form factors generated by the dynamics.
These first two contributions are independent on the parameter $\beta$ in the vacuum $\Sigma$. As we have demonstrated in the previous section, its value is not physical, as it can always redefined away as a phase of the techni-fermion spinors, thus in the following we will work in the simplest vacuum with $\beta = 0$.
The one loop of tops is given in Eq.~\ref{eq:Vtop}: after imposing the minimal conditions in Eq.~\ref{eq:zerotadpoles} to ensure the vanishing of the tadpoles for the triplets and the second doublet,
\beq
V_{\rm top} &=& - C_t f^4 \left( 8 Y_t^2 \sin^2 \theta + 2 \sqrt{2} Y_t^2 \sin (2\theta) \frac{h_1}{f} + \dots \right)\,.
\eeq

The total potential for $\theta$ reads:
\beq
V(\theta) =  - 8 C_t f^4 Y_t^2 \sin^2 \theta - C_g f^4 \frac{3 g^2 + {g'}^2}{2} \cos \theta^2 - 4 C_m f^3 (m_L + m_R) \cos \theta \,,
\eeq
which is minimised for
\beq \label{eq:min}
\left. \cos \theta \right|_{\rm min} = \frac{C_m (m_L + m_R)}{4 f C_t Y_t^2 \left( 1 - \frac{3 g^2 + {g'}^2}{16} \frac{C_g}{C_t Y_t^2}\right)}
\eeq
which is the same as one obtains in the minimal case SU(4)/Sp(4)~\cite{Galloway:2010bp,Cacciapaglia:2014uja,Katz:2005au}.

Expanding the potential at higher order allows to compute the masses of the pNGBs: general formulas for the mass terms can be found in Appendix~\ref{app:masses}. Using the above minimum condition to eliminate $C_m$ in favour of $\theta$, the mass of the Higgs-like state $h_1$, which does not mix with other pNGBs, is given by
\beq
m_{h_1}^2 = f^2 \sin^2 \theta \left( 2 C_t Y_t^2 - C_g \frac{3 g^2 + {g'}^2}{8} \right) = \frac{C_t}{4} m_{\rm top}^2 - \frac{C_g}{16} (2 m_W^2 + m_Z^2)\,.
\eeq
The other state that does not mix to other pNGBs is the pseudo-scalar singlet $s$, whose mass is given by
\beq
m_s^2 = \frac{m_{h_1}^2}{\sin^2 \theta}\,.
\eeq
Interestingly, the masses are the same as the ones obtained in the minimal SU(4)/Sp(4) case, however the coefficients $C_t$ and $C_g$, which depend on the underlying FCD, will differ.

\begin{figure}[tb]
\begin{center}
\includegraphics[width=6cm]{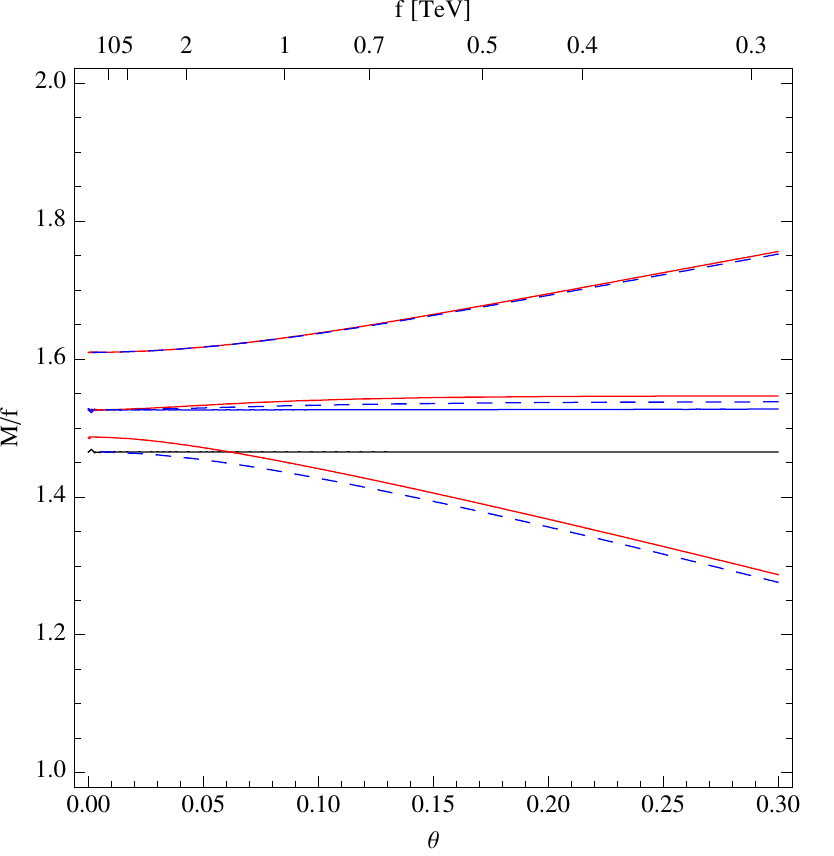}
\includegraphics[width=6cm]{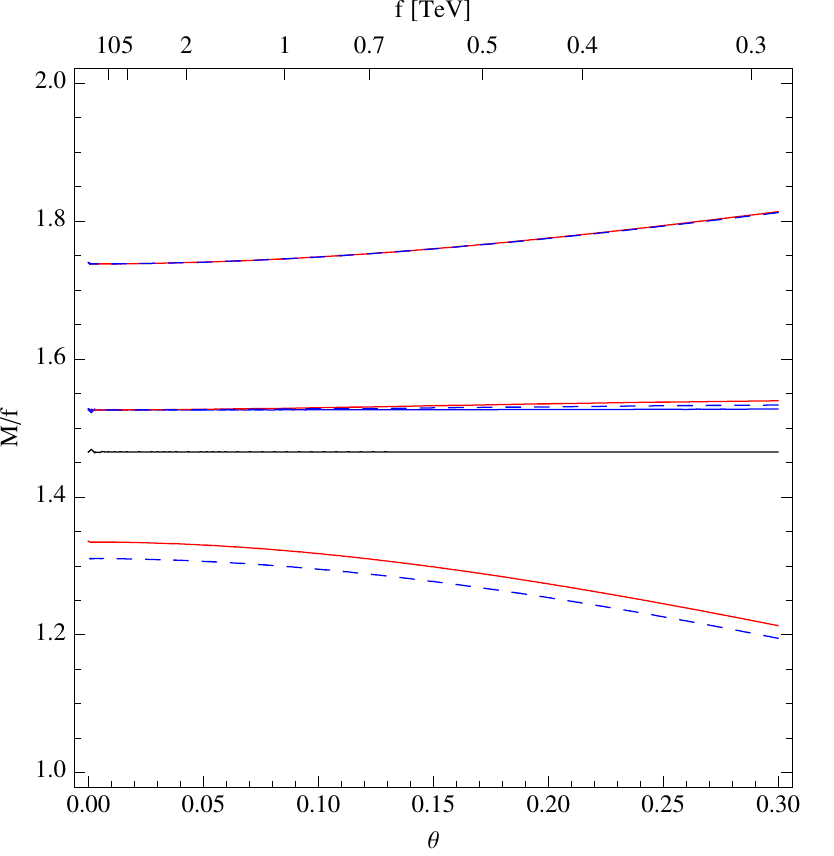}
\end{center}
\caption{Mass splitting of the pNGBs as a function of $\theta$ ($f$) for $C_g = 1$ and $\delta = 0$ (left), $\delta = 0.2$ (right). In solid black, the singlet $s$, in solid blue $A_0$, in red the charged states and in dashed blue the neutral pseudo-scalars.} \label{fig:spectrum}
\end{figure}

The spectrum of the other states is more complicated due to non-trivial mixings, generated by the top and gauge loop corrections. To have an approximate feeling of the behaviour, we can limit ourselves to the ``ETC'' Yukawa case, where $Y_0 = Y_t$ and $Y_T = 0$: in this case, $A_0$ decouples from the other neutral scalars, and its mass is
\beq
m_{A_0}^2 = 2 C_t Y_t^2 f^2 + C_g \frac{g^2-{g'}^2}{8} f^2 \sin^2 \theta \sim  \frac{m_{h_1}^2}{\sin^2 \theta} + \mathcal{O} (g^2, {g'}^2)\,.
\eeq
The remaining 3 charged and neutral states mix with each other: neglecting the smaller gauge corrections, the mass matrices, in the bases $\{ h_2, \Delta_0, N_0\}$ and $\{ H^\pm, \Delta^\pm, N^\pm\}$, are given by
\beq
M^2 = 2 C_t Y_t^2 f^2 \left( \begin{array}{ccc}
1 & \pm \frac{\sin \theta}{\sqrt{2}} & \frac{\sin \theta}{\sqrt{2}} \\
\pm \frac{\sin \theta}{\sqrt{2}}  & 1 + \delta \cos \theta & 0 \\
 \frac{\sin \theta}{\sqrt{2}} & 0 & 1-\delta \cos \theta
 \end{array} \right)\,,
 \eeq
 where the $+$ ($-$) is for the neutral (charged) masses. The mass eigenstates, which are equal for the two matrices, up to gauge corrections, are
 \beq \label{eq:simplyM}
 m_{1,2,3}^2 = 2 C_t Y_t^2 f^2 \times \left\{ \begin{array}{l}
 1 - \sqrt{\sin^2 \theta + \delta^2 \cos^2 \theta} \\
 1 \\
  1 + \sqrt{\sin^2 \theta + \delta^2 \cos^2 \theta}
  \end{array} \right.
\eeq
We see, therefore, that all the additional states have a mass of order $m_{h_1}/\sin \theta \sim f$.
The degeneracies among such states are thus removed by gauge corrections.
A numerical study of the spectrum is shown in Figure~\ref{fig:spectrum}, where we plot the ratio between the pNGB masses and the scale $f$ as a function of $\theta$, in the case of ``ETC'' Yukawas.
In the numerical examples, we use $m_W$, $m_{\rm top}$ and $m_h = 125$ GeV as inputs to fix the values of $Y_t$, $C_t$ and the relation between $f$ and $\theta$.

\subsection{Bounds from the Higgs couplings and EWPTs}

Like any other model of strong dynamics, our mode suffers from corrections to electroweak precision tests (EWPTs), that can be conveniently expressed in terms of the $S$ and $T$ parameters~\cite{Peskin:1990zt}.  These two parameters are sufficient to characterise precision constraints in this model: in fact, assuming that flavour physics is well reproduced by the UV completion generating the four fermion operators, non-universal corrections to the gauge couplings (like the $Zb\bar{b}$ coupling) are avoided. Furthermore, large contributions to LEP2 observables~\cite{Barbieri:2004qk,Cacciapaglia:2006pk} can be assumed small because axial/vector resonances should appear at a sufficiently high energy (see discussion in Section~\ref{sec:lattice}).

To estimate the impact on $S$ and $T$, we will follow the same procedure as in~\cite{Arbey:2015exa}: we divide the corrections in 3 contributions
\beq
\Delta S = \Delta S_{\rm Higgs} + \Delta S_{pNGB} + \Delta S_{FCD}\,,
\eeq
and similarly for $T$, where the first term, $\Delta S_{\rm Higgs}$ comes from the modification of the Higgs couplings and is Log--sensitive to the cut-off of the effective field theory $\Lambda_{FCD} \sim 4 \pi f$, the second, $\Delta S_{pNGB}$ contains the loop corrections from the additional light pNGBs, and finally $\Delta S_{FCD}$ contains the UV contribution of the strong dynamics. It should be noted, however, that the 3 contributions are not really independent, as both the Higgs-like state and the other pNGBs are part of the fundamental dynamics~\cite{Foadi:2012ga}. In fact, the scheme we use is to separate out the contribution of the light degrees of freedom from the heavy ones: thus, $\Delta S_{FCD}$ encodes, schematically, loops of the heavier bound states, like the axial/vectors in vector meson dominance.
The contribution of the Higgs can be estimated by rescaling the scalar loop and subtracting the contribution of the SM Higgs
\beq
\Delta S_{\rm Higgs} = \frac{1-\kappa_V^2}{6 \pi} \ln \frac{\Lambda_{FCD}}{m_h}\,, 
\quad \Delta T_{\rm Higgs} = - \frac{3 (1-\kappa_V^2)}{8 \pi \cos^2 \theta_W} \ln \frac{\Lambda_{FCD}}{m_h}\,,
\eeq
where $m_h = 125$ GeV is the measured Higgs mass, and $\kappa_V = \cos \theta$ is the ratio of the coupling of the Higgs-like state $h_1$ to SM gauge boson over the SM prediction.
This contribution is common to most composite Higgs models~\cite{Barbieri:2007bh}. Note also that the cut-off $\Lambda_{FCD}$ is close in value to the masses of the spin-1 states, so that it marks the separation of the low energy contribution of the light scalars from the contribution of the heavier resonances.
The second terms are generated by loops of the additional pNGBs: the second doublet and the triplets.
In the limit where all the masses are degenerate, we find
\beq \label{eq:ST}
\Delta S_{pNGB} = - \frac{\sin^2 \theta}{4 \pi}\,, \quad \Delta T_{pNGB} = \frac{\sin^2 \theta}{8 \pi \sin^2 \theta_W} \frac{m_{H\pm}^2 - m_{A_0}^2}{m_W^2} \ln \frac{\Lambda_{FCD}}{m_{pNGB}} \sim 0\,,
\eeq
where the $T$ parameter is proportional to the mass splitting between the charged Higgs and the CP-odd neutral one in the doublet, and is therefore small (smaller than the contribution of the Higgs).
The last contribution can be approximate by the contribution of loops of techni-fermions (thus diagrammatically close to the contribution of the spin-1 resonances), and reads~\cite{Peskin:1990zt,Maekawa:1994yd}
\beq
\Delta S_{FCD} = \frac{\sin^2 \theta}{3 \pi} N\,, \qquad \Delta T_{FCD} \sim 0\,,
\eeq
where $N$ is the number of FCD colours, and $T$ vanishes as the dynamics is approximately custodial invariant by construction.
Note that the result for $S$ and $T$ are nothing but a rough estimate due to the intrinsic non-perturbativity of the model we are studying, and one would have to rely on Lattice results for a more precise calculation (once the proper identification of contribution has been done~\cite{Foadi:2012ga}).

A more recent measurement that poses relevant constraints on the value of $\theta$ followed the discovery of the Higgs boson with the determination of its couplings to SM particles~\cite{Khachatryan:2014jba,ATLAS:2015bea}.
The simplest way to analyse the Higgs couplings is to parametrise the ratio of the couplings on the SM prediction, and compare this to the experimental results. We will use the parametrisation proposed in~\cite{Cacciapaglia:2009ky}, where the contribution of loops has been separated out from the modification of tree-level couplings.
In our model, 4 parameters are relevant:
\beq
\kappa_V = \cos \theta\,, \quad \kappa_f = \cos \theta\,, \quad \kappa_{\gamma \gamma} = - \frac{3 \tan^2 \theta}{16}\,, \quad \kappa_{gg} = 0\,.
\eeq
The first two contain the tree level modification to the couplings to massive gauge bosons, $WW$ and $ZZ$ which are equal due to the custodial invariance, and the modification to fermions, which are also assumed to be universal and equal to the one for the top.
The last two contain the loop contributions of new states to the couplings to photons and gluons: the coupling to photon is corrected by the contribution of loops of the charged component of the second doublet and the triplets. For the calculation, we used the masses in the simplified case as in Eq.~\ref{eq:simplyM}.
For the fitting procedure, we follow Ref.s~\cite{Cacciapaglia:2012wb,Flament:2015wra}.

\begin{figure}[tb]
\begin{center}
\includegraphics[width=8cm]{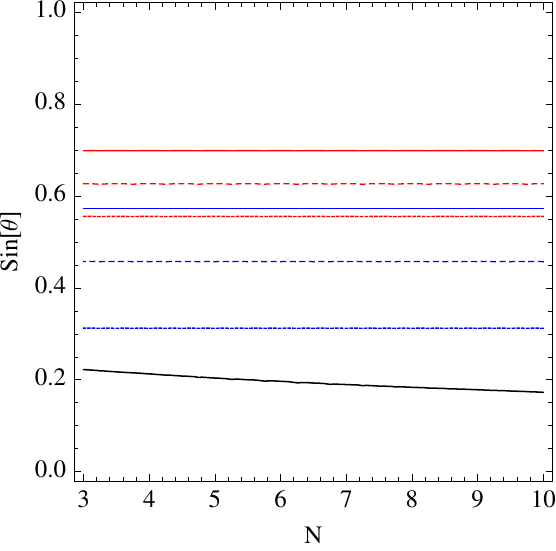}
\end{center}
\caption{Upper bound on $\sin \theta$ from EWPTs on $S$ and $T$ as a function of the number of FCD colours $N$. For comparison, we show the upper bounds derived from the Higgs coupling measurements at CMS (red) and ATLAS (blue), where the lines correspond to 1$\sigma$ (dotted), 2$\sigma$ (dashed) and 3$\sigma$ (solid).} \label{fig:bounds}
\end{figure}

The numerical results are shown in Figure~\ref{fig:bounds}. In black, we show the upper bound on $\sin \theta$ as a function of the number of FCD colours $N$: the plot shows a mild dependence on the number of colours, while the constraints is set around $\sin \theta \lesssim 0.2$. To be more specific, for $N = 3$ we find $\sin \theta < 0.22$, while for $N=4$, we obtain $\sin \theta < 0.21$. In the same figure we also show the constraints from the Higgs coupling measurements, which are independent on the number of FCD colours. The constraints are the same as we found in the minimal case~\cite{Arbey:2015exa}, except for the contribution of the charged pNGBs to the di-photon decays: numerically we find that at 3$\sigma$ CMS imposes a bound $\sin \theta < 0.64$, while ATLAS requires $\sin \theta < 0.57$. We do not attempt to combine the two experiments, as this would require a thoroughly understanding of the systematic uncertainties.
The bounds from the Higgs measurements are milder that the constraint from EWPTs, however the improvement in the measurements at LHC Run--II will certainly increase their relevance.

\subsection{The bottom mass, and flavour alignment}

The bottom mass can be generated in a similar way as the top one, by adding 4-fermion interactions that, in the low energy effective theory, generate terms similar to Eq.~\ref{eq:Yuk0}:
\beq \label{eq:Yukb0}
\mathcal{L}_{\rm Yuk,b} = - f\, (\bar{\chi}_{q_L} \bar{\eta}_{b_R})^\alpha \left[ \mbox{Tr} [P_{b1,\alpha} (y_{b1} \Sigma + y_{b2} \Sigma^\dagger)] - (i\sigma_2)_{\alpha \beta} \mbox{Tr} [P_{b2}^\beta (y_{b3} \Sigma + y_{b4} \Sigma^\dagger)] \right] + h.c.
\eeq
where the projectors are defined in terms of the top ones as $P_{b1,\alpha} = (P_2^\alpha)^\dagger$ and $P_{b2}^\alpha = (P_{1,\alpha})^\dagger$.
After defining the combination of pre-Yukawas
\beq
Y_b = \frac{y_{b1} - y_{b2} - (y_{b3} - y_{b4})}{2 \sqrt{2}}\,, &  & Y_{bD} = \frac{y_{b1} - y_{b2} + (y_{b3} - y_{tb})}{2 \sqrt{2}}\,, \nonumber \\
Y_{bT} = \frac{y_{b1} + y_{b2} + (y_{b3} + y_{b4})}{2 \sqrt{2}}\,, &  & Y_{b0} =  \frac{y_{b1} + y_{b2} - (y_{b3} + y_{b4})}{2 \sqrt{2}}\,,
\eeq
expanding Eq.~\ref{eq:Yukb0} to linear order in the pNGB fields yields, for $\beta = 0$,
\beq \label{eq:Yukb1}
\mathcal{L}_{\rm Yuk,b} &=& - \left[  Y_b  v_{SM} + Y_b \cos \theta\ h_1 + i Y_{bD}\ h_2 - Y_{bD} \cos \theta\ A_0 - i \frac{Y_{bT}}{\sqrt{2}} \sin \theta\ (N_0 + \Delta_0)\right] (\bar{\chi}_{b_L} \bar{\eta}_{b_R}) \nonumber \\
&& - \left[ i \sqrt{2} Y_{bD} \cos \theta\ H^- + i Y_{bT} \sin \theta\ (N^- + \Delta^-) \right] (\bar{\chi}_{t_L} \bar{\eta}_{b_R}) + h.c.
\eeq
which is very similar to Eq.~\ref{eq:Yuk1}, up to the signs of the couplings of $A_0$, $N_0$, $\Delta_0$ and $H^\pm$.
The contribution to the potential of the pNGBs also resembles the top one in Eq.~\ref{eq:Vtop0}, up to signs:
\begin{multline} \label{eq:Vbot0}
V_b = - f^4 C_t \left(8 |Y_b|^2 \sin^2 \theta + 2 \sqrt{2} |Y_b|^2 \sin (2 \theta) \frac{h_1}{f} - i\, 2 \sqrt{2} (Y_{bD}^\ast Y_b - Y_{bD} Y_b^\ast) \sin \theta \frac{h_2}{f} \right.  \\
 \left. - \sqrt{2} (Y_{bD}^\ast Y_b + Y_{bD} Y_b^\ast) \sin (2\theta) \frac{A_0}{f} + i\, 2 (Y_{bT}^\ast Y_b - Y_{bT} Y_b^\ast) \sin^2 \theta \frac{N_0 + \Delta_0}{f} + \dots \right)\,.
\end{multline}
We expect the coefficient $C_t$ generated by the dynamics to be the same as for the top, as the structure of the operator under the FCD is the same.
Remarkably, the two terms with different sign are the tadpoles for $A_0$ and for the triplets, which would violate custodial symmetry.  This fact becomes clear when looking at Eq.s~\ref{eq:Yuk0} and \ref{eq:Yukb0}: assembling $t_R$ and $b_R$ into an SU(2)$_R$ doublet would in fact require that $y_{ti} = y_{bi}$ for $i = 1,\dots 4$, thus any violation of custodial invariance should be proportional to the difference of pre-Yukawas. In fact, the tadpole for $A_0$ is proportional to
\beq
(Y_D^\ast Y_t + Y_D Y_T^\ast) - (Y_{bD}^\ast Y_b + Y_{bD} Y_b^\ast) =  \mbox{Re} (\delta Y_D) (Y_t + Y_b) + \delta Y_f \mbox{Re} (Y_D + Y_{bD})\,,
\eeq
where $\delta Y_D = Y_D - Y_{bD}$ and $\delta Y_f = Y_t - Y_b$, and we have assumed real $Y_t$ and $Y_b$.
A similar analysis can be done for the triplet tadpole.
As it is not possible to set all pre-Yukawa differences to zero (we know that $Y_t \neq Y_b$), the only way to ensure a custodial invariant vacuum is to have $\mbox{Re}(Y_D) = \mbox{Re} (Y_{bD}) = 0$ and $\mbox{Im}(Y_T) = \mbox{Im} (Y_{bT}) = 0$: these conditions are automatically ensured in the case of ``BTC'' interactions.

The custodial invariant $h_2$ tadpole, on the other hand, is connected to the presence of $\beta$ in the vacuum. As we already discussed, $\beta$ in the vacuum can be removed by the transformation in Eq.~\ref{eq:Omegabeta}, which corresponds to the redefinition of an unphysical phase in the techni-fermion fields. The procedure to follow, therefore, is the following: we minimise the potential by ensuring the vanishing of the $h_2$ tadpole, thus determining $\beta$ as a function of the phases in the top and bottom Yukawas; we then use $\Omega_\beta$ to set $\beta = 0$ on the vacuum, and at the same time changing the phases of the top and bottom Yukawas ($Y_D$ and $Y_{bD}$), without however loss of generality as we are simply fixing an unphysical phase in the FCD. This reasoning shows that one can always work in the $\beta = 0$ vacuum, and think of the vanishing of the $h_2$ tadpole as of the fixing of an arbitrary phase.
It is interesting that in the ``BTC'' case, the phases of $Y_D$ and $Y_{bD}$ are aligned, so that one can make both real with the same phase redefinition.

The masses for the light generations, and flavour mixing, can also be added to the model by promoting the pre-Yukawas $y_{ti}$ and $y_{bi}$ to matrices in the SM flavour space. In the most general set up, the model will however suffer from large flavour changing neutral currents (FCNCs) generated by the  couplings of the second doublet and the triplets. This problem can be avoided if the combinations of Yukawas $Y_t$, $Y_D$ and $Y_T$ (and similarly for the down-type quarks) can be simultaneously aligned.
The FCNC-free scenario would therefore correspond to pre-Yukawa couplings which are all proportional to the same flavour matrix:
\beq
y_{ti} = \lambda_t^{ab} y_i\,, \quad y_{bi} = \lambda_b^{ab} y_i\,,
\eeq
where the pre-Yukawas $y_i$ parametrise universal couplings of the mediators to the techni-fermions, while the $\lambda$ matrices contain all the information about the quark masses and flavour mixing. This scenario corresponds to a minimal flavour violation setting, and it naturally arises in ``BTC'' frameworks.
Another possibility would be to generate the top (and bottom) masses via partial compositeness, while the light quarks are generated by 4-fermion interactions, thus potentially suppressing FCNCs~\cite{Cacciapaglia:2015dsa}.

\section{Discrete symmetries and Dark Matter candidates}
\label{sec:DM}
\setcounter{equation}{0}
\setcounter{footnote}{0}

Besides the Higgs-like scalar $h_1$ and the eaten Goldstone bosons, the model contains 11 additional pNGBs: the chiral Lagrangian one can write down respecting the symmetry breaking patters is invariant under a parity changing sign to all pNGBs, thus they only appear in bilinear couplings.
This property is however violated by the explicit symmetry breaking terms: we have seen this in the loop-induced potential, which generates mixing between scalars, and the couplings to the top quarks.
In order to understand if any of the additional pNGB may be stable, it is useful to think in terms of multiplets of the electroweak symmetry, as different states within a multiplet are always connected by gauge interactions. Thus, in the limit $\theta = 0$, the model contains a second doublet $H_2$, a SU(2)$_L$ triplet $\Delta$, a SU(2)$_R$ triplet $N$ (consisting on a charged and a neutral singlet), and a singlet $s$. To identify a Dark Matter candidate we need to establish both the mixing patterns among the multiplets, and their direct couplings to SM states. 

The mass mixing structures we found in Appendix~\ref{app:masses} can be summarised as follows:
\begin{itemize}
\item[-] gauge interactions mix the two triplets, $\Delta$ and $N$;
\item[-] top Yukawa couplings mix the doublet with the triplets, with a coupling proportional to $Y_0$;
\item[-] top Yukawa couplings mix the two triplets with coupling proportional to $Y_T$.
\end{itemize}
We see already that the singlet $s$ does not mix with the others states. While gauge interactions cannot be turned off, the Yukawa couplings involved in the mixing may be zero depending on their origin, and we will be particularly interested in $Y_0$, which generates mixing between the doublet and the triplets.

Regarding possible decay channels, there are two terms in the lowest order effective Lagrangian that generate couplings of a single pNGB to SM states: one is due to the couplings to the tops, and another to the Wess-Zumino-Witten (WZW)~\cite{Wess:1971yu,Witten:1983tw} anomaly. We have already seen in the previous section that, in a custodial preserving vacuum, only the triplets are allowed direct couplings to tops via a combination of Yukawas $Y_T$.
The WZW term, on the other hand, is generated by a triangle loop of techni-fermions and it contains potential couplings of the pNGBs to EW gauge bosons.  The pNGBs can be associated to the following current
\beq
J_5^{\mu a} = \bar{\psi} \gamma^\mu \Omega T^a \Omega^\dagger \frac{1+\gamma^5}{2} \psi -  \bar{\psi} \gamma^\mu \Omega^\dagger T^a \Omega \frac{1-\gamma^5}{2} \psi \,,
\eeq
where $\Omega = \Omega (\theta, 0)$ in Eq.~\ref{eq:Omega}. Following the results in Ref.s~\cite{Witten:1983tw,Kaymakcalan:1983qq}, the result of the triangle anomaly can be expressed as
\begin{multline}
\Gamma_{WZW} \sim \left\{ a_1\ \mbox{Tr} \left[ T^a T^b (\Omega \Pi \Omega^\dagger + \Omega^\dagger \Pi \Omega) \right] + \right. \\
\left. a_2\ \mbox{Tr} \left[ T^a (\Omega \Pi \Omega T^b \Omega^\dagger \Omega^\dagger + \Omega^\dagger \Pi \Omega^\dagger T^b \Omega \Omega ) \right] \right\} V^a V^b\,,
\end{multline}
where $T^{a,b}$ are now the gauged generators of the global symmetry SU(4)$^2$: the first term of the above expression can be understood as a triangle anomaly of the current $J_5$, while the second term derives from a box diagram.
As a result, we can extract the following couplings:
\beq
\mathcal{L}_{WZW} \sim k\ c_\theta\, s (g^2 W_{\mu \nu} \tilde{W}^{\mu \nu} - {g'}^2 B_{\mu \nu} \tilde{B}^{\mu \nu})\,,
\eeq
where $k$ is a numerical factor. Interestingly, the WZW anomaly only involves the singlet $s$, and its couplings are similar to the ones in the minimal SU(4)/Sp(4) case~\cite{Galloway:2010bp,Arbey:2015exa}: in particular, no coupling to two photons is generated. This result shows that $s$ cannot play the role of Dark Matter.

The only pNGBs that may play the role of Dark Matter are therefore the triplets and the second doublet.
Their mixing and decays are ruled by the top Yukawa couplings, as discussed above: $Y_0$ induces a mass mixing between the doublet and the triplets, $Y_T$ induces decays of the triplets directly to tops.
The situation can be summarised as follows:
\begin{center}
\begin{tabular}{c|c|c|}
DM candidates   &   $Y_T = 0$ & $Y_T \neq 0$  \\
\hline
$Y_0=0$ &  $H_2$ and $\Delta$--$N$  & $H_2$  \\
\hline
$Y_0 \neq 0$  & mixed   &  no DM \\
\hline
\end{tabular}
\end{center}
This analysis, based on the lowest order Lagrangian, is not conclusive as additional mixing/decays may be generated by higher order terms in the Lagrangian: we thus need to identify a symmetry that protects the DM candidate.

As the techni-fermions are vector-like with respect to the FCD gauge and the SM ones, the strong sector will be invariant under P and C separately, which act on the pNGB matrix and gauge bosons as
\beq
P(\Sigma) = \Sigma^\dagger \,, \quad P(A_\mu) = - (-1)^{\delta^{0\mu}} A_\mu\,, \qquad C(\Sigma) = \Sigma^T\,, \quad C(A_\mu) = - A_\mu^T\,.
\eeq
The vacuum, however, is not invariant under C nor P: it is invariant under CP only for $\beta=0$.
We identified 2 symmetries that act as parities on the pNGB fields:
\begin{itemize}

\item[- A:] parity $P$ combined with an SU(4) transformation, acting as
\beq
\Sigma \to P_A \Sigma^\dagger P_A^\dagger \,, \quad A_\mu \to - (-1)^{\delta^{0\mu}} A_\mu\,, \quad 
\mbox{with}\;\;
P_A = \left( \begin{array}{cc}
1 & 0 \\ 0 & -1 \end{array}\right)\,.
\eeq
Under this symmetry, $\Pi \to - P_A \Pi P_A^\dagger$, thus $s$ and the triplets $\Delta$ and $N$ are odd. The top Yukawas break the symmetry unless the following relation between the couplings is imposed~\footnote{This relations are only possible for ``BTC'' pre-Yukawas, but not for ``ETC'' ones.}:
\beq
y_{t2} = - y_{t1}\,, \quad y_{t4} = - y_{t3}\,, \qquad \Rightarrow \;\; Y_T = Y_0 = 0\,.
\eeq
This symmetry, however, is broken in the present model: besides the gauge interactions of the SM fermions, that violate P, the WZW term is allowed by this symmetry as it couples odd scalars to a P-odd combination of vector bosons. In principle, a WZW term is allowed for both the singlet and the triplets.

\item[- B:] a second symmetry we identified acts as charge conjugation plus a global SU(4):
\beq \label{eq:parityB}
\Sigma \to P_B \Sigma^T P_B^\dagger \,, \quad A_\mu \to - P_B A^T_\mu  P_B^{\dagger} = A_\mu\,, \quad 
\mbox{with}\;\;
P_B = \left( \begin{array}{cc}
\sigma_2 & 0 \\ 0 & -\sigma_2 \end{array}\right)\,.
\eeq
The vacuum $\Sigma_1$, however is only invariant if $\beta = 0$. The pNGBs transform as $\Pi \to P_B \Pi^T P_B^\dagger$, thus we find that the triplets $\Delta$ and $N$ and the second doublet $H_2$ are odd. In this case, the gauge interactions of the SM are also invariant. A condition is nevertheless needed on the top pre-Yukawas:
\beq
y_{t3} = - y_{t1}\,, \quad y_{t4} = - y_{t2}\,, \qquad \Rightarrow \;\; Y_D = Y_T = 0\,.
\eeq
Note that, as the dynamics respects this symmetry, a WZW term for the triplets is forbidden in general. Also, a model invariant under this symmetry has automatically a custodial invariant vacuum.

\end{itemize}

From the above analysis we can conclude that the only viable Dark Matter candidates are the second doublet and the triplets, in models where the symmetry B is preserved (i.e. $Y_T = Y_D = 0$). Note that this condition can be satisfied by BTC Yukawa couplings, with $y_{t2} = - y_{t1}^\ast$. Furthermore, for an imaginary $Y_0$, one needs to identify the symmetry ``GP''  to the ordinary CP. We also checked that the full WZW term is invariant under the symmetry B, so that no violation is present at any order in the pNGB field expansion.

\section{Spectrum of resonances, and Lattice results}
\label{sec:lattice}
\setcounter{equation}{0}
\setcounter{footnote}{0}

Insofar we have focused on the physics of the light scalar degrees of freedom of the theory, i.e. the pNGBs, however the model also contains massive composite states of other spins.
We are particularly interested in Baryonic bound states, as they carry TB number and are therefore stable and potential candidates for (asymmetric) Dark Matter. The properties of such states depend crucially on the number of FCD colours in SU(N)$_{\rm FCD}$, as the bound state will be made of N techni-fermions: if N is odd, the bound state will be a fermion, while for even N it will be a boson. We will focus here, for concreteness, on the smallest numbers of FCD colours, i.e. $N=3$ and $N=4$.

For $N=3$, the baryons are made of 3 techni-fermions, thus they belong to the following representations of the flavour group SU(4):
\beq \label{eq:444}
{\bf 4} \otimes {\bf 4} \otimes {\bf 4} = {\tiny{\yng(1,1,1)}} \oplus 2 \times {\tiny{\yng(2,1)}} \oplus {\tiny{\yng(3)}} = {\bf \bar{4}} \oplus 2 \times {\bf 20} \oplus {\bf 20''}\,.
\eeq
It should be recalled here that the FCD colour indices are fully anti-symmetric, thus the wave function in terms of the flavour indices, spin and orbital momentum should be overall symmetric.
To identify the ground state, i.e. states that have zero orbital momentum, it is useful to include the spin indices into the flavour ones: each techni-fermion is thus doubled into two states with spin up and spin down respectively, and the global symmetry is thus extended to SU(8).
The ground state, which has no orbital momentum, must therefore be fully symmetric in the SU(8) space, and it thus belongs to the 3-index symmetric representation ${\bf 120}_{\rm SU(8)}$. Under spin and SU(4), it decomposes into:
\beq
{\bf 120}_{\rm SU(8)} = \mbox{spin-1/2} \times {\bf 20} \oplus \mbox{spin-3/2} \times {\bf 20''}\,.
\eeq
The other states in the decomposition in Eq.~\ref{eq:444} must therefore carry some orbital momentum, and they belong to heavier excited states. The spin-1/2 bound states decompose under the custodial SU(2)$_L\times$SU(2)$_R$ as:
\beq
{\bf 20} = (3,2) + (2,3) + (2,1) + (2,1) + (1,2) + (1,2)\,.
\eeq
All the states in this multiplet have semi-integer electric charge, $\pm 1/2$ and $\pm 3/2$: in order to avoid the strong bounds on stable non-integer charge states~\cite{Langacker:2011db}, we can partly charge the TB number, so that the ordinary hypercharge is generated by $T_R^3 + TB$, without affecting the properties of the Higgs-like states and of the pNGBs, which do not carry TB number. Notice that this partial gauging does not break a global TB, which remains a conserved number. Assigning gauged TB equal to $+1/2$ or $-1/2$, all the stable spin-1/2 states will have integer charges, with the neutral components that may play the role of Dark Matter.

In the case $N=4$, the baryons are made of 4 techni-fermions and are therefore bosons.
They decompose under SU(4) as:
\beq \label{eq:4444}
{\bf 4} \otimes {\bf 4} \otimes {\bf 4} \otimes {\bf 4} &=& {\tiny{\yng(1,1,1,1)}} \oplus 3 \times {\tiny{\yng(2,1,1)}} \oplus 3 \times {\tiny{\yng(3,1)}} \oplus 2 \times {\tiny{\yng(2,2)}} \oplus {\tiny{\yng(4)}} \nonumber \\
&=& {\bf 1} \oplus 3 \times {\bf 15} \oplus 3 \times {\bf 45} \oplus 2 \times {\bf 20'} \oplus {\bf 35}\,.
\eeq
To identify the ground state, we follow the same procedure as above: the 4-index symmetric representation of SU(8) is a ${\bf 330}_{\rm SU(8)}$, which decomposes as
\beq
{\bf 330}_{\rm SU(8)} = \mbox{spin-0} \times {\bf 20'} \oplus \mbox{spin-1}\times {\bf 45} \oplus \mbox{spin-2} \times {\bf 35}\,.
\eeq
The lowest spin scalar baryons, thus, belong to a ${\bf 20'}$ rep of SU(4), which decomposes under the custodial symmetry as
\beq
{\bf 20'} = (3,3) + (2,2) + (2,2) + (1,1) + (1,1)\,.
\eeq
In this case, all the states have integer charges and the multiplet contains neutral states which are candidates for Dark Matter.\\

The model also contains spin-1 resonances, common to any model of compositeness. Like in QCD, the lightest resonances consist on a set of vector (CP-even) states and a set of axial (CP-odd) states, associated respectively to the fermionic currents:
\beq
\rho^\mu = \langle \bar{\psi} \gamma^\mu \psi \rangle\,, \quad a^\mu = \langle \bar{\psi} \gamma^\mu \gamma^5 \psi \rangle\,,
\eeq
where $\psi$ are the techni-fermion Dirac spinors.
Both vector and axial mesons transform as the adjoint of the unbroken SU(4) group, thus they transform under the SU(2)$_L \times$ SU(2)$_R$ subgroup like the pNGBs:
\beq
\rho^\mu\,, \;\; a^\mu = (2,2) + (2,2) + (3,1) + (1,3) + (1,1)\,.
\eeq
The phenomenology of the triplets is similar to the one of vector resonances in minimal models~\cite{Pappadopulo:2014qza,Becciolini:2014eba}: as they have the same quantum numbers of the SM gauge bosons, they will mix with them in the effective Lagrangian, and thus acquire a direct coupling to the SM fermions. They will therefore be produced at the LHC in Drell-Yan, and decay either into a pair of fermions or into a pair of gauge bosons.
On the other hand, the properties of the doublets can be quite novel: due to their quantum numbers, they cannot couple directly to the SM fermions. Their only couplings may therefore involve the additional pNGBs present in the model. We postpone a detailed study of their couplings to a further study. 
In cases where the model has a Dark Matter candidate, as detailed in Section~\ref{sec:DM}, some of the spin-1 resonances may be odd under the same parity stabilising the Dark Matter pNGB candidate. We verified that, under the parity B in Eq.~\ref{eq:parityB}, one of the doublets and the singlet vectors, together with the other doublet and the triplets of the axial states, are odd and therefore can only decay into a stable pNGB.

Lattice results~\cite{Hasenfratz:2009ea,Fodor:2009wk,Aoki:2012eq,Aoki:2013xza} are very useful in the study of the vector resonances due to the relative ease in extracting their masses from data.
In~\cite{Fodor:2009ff}, it is reported that the mass of the vectors in the case SU(3)$_{\rm FCD}$, in units of the pNGB decay constant is
\beq
\frac{M_\rho}{F_\pi} = 13 \pm 1\,, \quad \left( \left.\frac{M_\rho}{F_\pi}\right|_{\rm QCD} \sim 8 \right)\,;
\eeq
where we show, for comparison, the ration in QCD (with 3 flavours). Rescaling the value of the mass to the EW scale, $F_\pi = 246$ GeV, we find a mass $M_\rho \sim 3.2$ TeV in the TC limit (i.e. $\sin \theta = 1$).
In the pNGB Higgs limit, the mass should be multiplied by a factor $1/\sin \theta$, thus for $\sin \theta < 0.22$ one obtains $M_\rho > 14$ TeV.
These preliminary results on the vector masses, therefore, indicate that they are expected to be very heavy and beyond the reach of the LHC Run-II. 
They may however be accessible to a higher energy proton collider, like the proposed 100 TeV colliders.

\section{Conclusions and Outlook}
\label{sec:concl}
\setcounter{equation}{0}
\setcounter{footnote}{0}

Compositeness as a paradigm to explain the origin of the Higgs boson, discovered at the LHC, is still one of most appealing extensions of the Standard Model.  In this work, we pursued compositeness by defining a fundamental composite dynamics (FCD) based on a simple confining gauge group plus fermionic matter. This approach has the advantage of guiding the building of the low energy chiral Lagrangian, and it can be simulated on the Lattice in order to have non-perturbative predictions of the spectrum. The need for numerical prediction is in fact essential for studying the viability of such models {\it vis a vis} the results at the LHC.

The minimal model of FCD has a global symmetry breaking pattern SU(4)/Sp(4). Here we focus on a less minimal case based on the symmetry breaking SU(4)$\times$SU(4)/SU(4), which is the smallest symmetry of this kind that enjoys custodial symmetry. The underlying dynamics is provided by a gauged SU(N)$_{\rm FCD}$ with 4 Dirac techni-fermions in the fundamental representation. This theory is known to condense.
We construct the effective Lagrangian for the 15 pseudo-Nambu-Goldstone bosons, which transform, in the limit of unbroken symmetry, as 2 bi-doublets, one SU(2)$_L$ and one SU(2)$_R$ triplet (a $6$ of the custodial SO(4)) and one singlet.
The model has therefore two potential Higgs doublets: the alignment of the EW symmetry breaking vacuum along the two doublets, however, depends on the structure of the interactions generating the top mass.
We found that, adding only a mass for the top, the vacuum is aligned with one of the two doublets, thus effectively generating a composite inert 2HDM.
Interestingly, the custodial invariant direction on the second doublet corresponds to a phase in the vacuum, which can be associated with a global U(1) subgroup of the SU(2)$^2$ symmetry. One can therefore use this symmetry to always set the second doublet vacuum to zero, without affecting the physical properties of the model. This U(1) corresponds in the FCD to an unphysical phase redefinition of the techni-fermion fields.
We also determine the conditions on the Yukawa couplings that ensure a custodial invariant vacuum.

The model suffers from contributions to electroweak precision observables, mainly the $S$ parameter: we show that such contributions can be under control when the angle parametrising the alignment along the EW breaking direction is small. We found that values $\sin \theta \lesssim 0.2$ are still allowed. The measurements of the Higgs couplings also pose a constraint on the angle, which is however milder at present, $\sin \theta \lesssim 0.57 \div 0.64$.
These constraints are very similar to the ones obtained in the minimal model, thus showing that less minimal cases are equally likely to be realised.

The most interesting feature of non-minimal cases is that the additional pNGBs may be stable due to residual unbroken parities. Under certain conditions on the Yukawa couplings, we identified a symmetry that protects the second doublet and the two triplets.
This symmetry is exact, and it is preserved by all the explicit breaking we add and by the entire Wess-Zumino-Witten term: the Dark Matter candidate is therefore a component of the second inert doublets, which mixes with the two triplets.
Finally, we studied the spectrum of the heavier composite states: spin-1 vector and axial resonances and spin-1/2 (or spin-0) techni-baryons. The latter are stable due to a conserved techni-baryon number, and may thus play the role of an asymmetric Dark Matter. However, the masses of such states are expected to lie in the $\mathcal{O} (10)$ TeV range, thus they may only be explored directly at a 100 TeV collider.

The model we explored here is very similar to QCD with 4 flavours. In fact, the case SU(3) has already been studied on the lattice and confirmed to condense. It would be very interesting to further study this model on the Lattice to calculate the masses of the bound states, in particular the vectors and techni-baryon. The spectrum can be a precious guide in defining the search strategies at the LHC and at a future 100 TeV collider, and also allow us to study in detail the relic abundance of the stable techni-baryons.

The SM flavour physics of this model may be very interesting, as each Yukawa coupling is generated by 4 operators (typically four-fermion interactions). However, the only way to reliably study flavour physics is by defining a UV completion that generates the needed four-fermion interactions. Another possibility that we plan to explore is to extend the model in order to have techni-baryons that may mix linearly with the top quark (top partners).

\section*{Acknowledgments}

We thank Mads Frandsen for inspiring us to start this project, and the CP3-Origins crew for useful discussions. We also thank J.B. Flament for providing to us the Mathematica files with the Higgs coupling fit results.
M.T. also thanks the IPNL for the hospitality during the completion of this work, and Tsinghua Scholarship for Overseas Graduate Studies for financial support. 
G.C. acknowledges partial support from the Labex-LIO (Lyon Institute of Origins) under grant ANR-10-LABX-66 and FRAMA (FR3127, F\'ed\'eration de Recherche ``Andr\'e Marie Amp\`ere").

\appendix

\section{Couplings to gauge bosons}
\label{app:couplings}
\setcounter{equation}{0}
\setcounter{footnote}{0}

In the gauge basis, defining
\beq
\varphi_1 \overleftrightarrow{\partial_\mu} \varphi_2 = \varphi_1 (\partial_\mu \varphi_2) - (\partial_\mu \varphi_1) \varphi_2,
\eeq
the couplings of a single gauge boson to the pNGBs can be written as:
\beq
\mathcal{L}_{\rm A} &=& i g s_W A^\mu \left( H^- \overleftrightarrow{\partial_\mu} H^+ + N^- \overleftrightarrow{\partial_\mu} N^+ + \Delta^- \overleftrightarrow{\partial_\mu} \Delta^+\right) \,, \\
\mathcal{L}_{\rm Z} &=& \frac{i g}{2 c_W} Z^\mu \left( c_{2W}\  H^- \overleftrightarrow{\partial_\mu} H^+   + i c_\theta\ A_0 \overleftrightarrow{\partial_\mu} h_2 +  (c_{2W} - c_\theta)\ N^- \overleftrightarrow{\partial_\mu} N^+  \right. \nonumber \\
& & \left. + (c_{2W} + c_\theta)\ \Delta^- \overleftrightarrow{\partial_\mu} \Delta^+ \right)\,, \\
\mathcal{L}_{\rm W} &=& \frac{i g}{2} W^{\mu,-} \left( c_\theta\ h_2 \overleftrightarrow{\partial_\mu} H^+ - i  A_0 \overleftrightarrow{\partial_\mu} H^+ - 2 s_{\theta/2}^2\  N_0 \overleftrightarrow{\partial_\mu} N^+ \right. \nonumber \\
& & \left.- 2 c_{\theta/2}^2\ \Delta_0 \overleftrightarrow{\partial_\mu} \Delta^+  \right) + h.c.
\eeq
where $c_W = \cos \theta_W$, $s_W = \sin \theta_W$, $c_{2W} = \cos 2 \theta_W$ and $c_\theta = \cos \theta$.

The couplings of 2 gauge bosons with 2 charged scalars can be written as:
\beq
\mathcal{L}_{\rm AA\pm} &=& g^2 s_W^2\ A_\mu A^\mu \left(  H^+ H^- + N^+ N^- + \Delta^+ \Delta^-  \right)\,, \\
\mathcal{L}_{\rm AZ\pm} &=& \frac{g^2 s_W}{c_W}\  A_\mu Z^\mu \left(  c_{2W}\ H^+ H^- + (c_{2W} - c_\theta)\ N^+ N^- + (c_{2W} + c_\theta)\ \Delta^+ \Delta^-  \right)\,, \\
\mathcal{L}_{\rm ZZ\pm} &=& \frac{g^2}{8 c_W^2} Z_\mu Z^\mu \left( 2 c_{2W}^2\ H^+ H^- + (3 c_\theta^2 - 4 c_{2W} c_\theta + c_{4W})\ N^+ N^- + \right. \nonumber \\
 & & \left. + (3 c_\theta^2 + 4 c_{2W} c_\theta + c_{4W})\  \Delta^+ \Delta^- -  s_\theta^2 (\Delta^+ N^- + N^+ \Delta^-) \right)\,, \\
 \mathcal{L}_{\rm WW\pm} &=& \frac{g^2}{2}\ W^{\mu,+} W_\mu^- \left(  c_\theta^2\ H^+ H^- - 2 s_{\theta/2}^2 c_\theta\  N^+ N^- + 2 c_{\theta/2}^2 c_\theta \ \Delta^+ \Delta^-   \right)\,, \\
 \mathcal{L}_{\rm WW++} &=&\frac{g^2}{4}\ W^{\mu,-} W_\mu^- \left(  s_\theta^2\ H^+ H^+  -2 (c_{\theta/2}^2\ \Delta^+  - s_{\theta/2}^2\ N^+)^2   \right) + h.c.\,;
\eeq

The quadrilinear couplings with neutral scalars are:
\beq
\mathcal{L}_{\rm ZZ0} &=& \frac{g^2 }{8 c_W^2}\ Z^\mu Z_\mu \left(  c_{2\theta}\ h_1^2 + c_\theta^2\  h_2^2  + c_{2\theta}\ A_0^2 - \frac{1}{2} s_\theta^2\ (N_0 - \Delta_0)^2  - s_\theta^2\ s^2 \right)\,, \\
\mathcal{L}_{\rm WW0} &=& \frac{g^2}{4}\ W^{\mu,+} W_\mu^- \left(  c_{2\theta}\ h_1^2 + c_\theta^2\  h_2^2  + A_0^2 + s_{\theta/2}^2 (1-3 c_\theta) \ N_0^2 \right. \nonumber \\
&& \left.  + c_{\theta/2}^2 (1+3 c_\theta)\ \Delta_0^2 - s_\theta^2\ N_0 \Delta_0  - s_\theta^2\ s^2\right)\,.
\eeq

Finally, for the charged currents:
\beq
\mathcal{L}_{\rm WA} &=& \frac{g^2 s_W}{2}\ A^\mu W_\mu^- \left(  c_\theta\ H^+ h_2  - i\ H^+ A_0 - 2 s_{\theta/2}^2\  N^+ N_0 - 2 c_{\theta/2}^2\ \Delta^+ \Delta_0  \right) + h.c.\,, \\
\mathcal{L}_{\rm WZ} &=& \frac{g^2}{2 c_W} \ Z^\mu W^-_\mu \left( - c_\theta s_W^2 H^+ h_2 - i (c_W^2 - c_\theta^2)\ H^+ A_0  - s_{\theta/2}^2 (c_{2W} - c_\theta)\ N^+ N_0   \right. \nonumber \\
& & \left.  - c_{\theta/2}^2 (c_{2W} + c_\theta)\ \Delta^+ \Delta_0 + \frac{1}{2} s_\theta^2\ (N^+ \Delta_0 + \Delta^+ N_0)  \right) + h.c. \,.
\eeq

\section{ Most general vacuum structure.}
\label{app:phases}
\setcounter{equation}{0}
\setcounter{footnote}{0}

\subsection{Custodial invariant vacua}

To better understand the conditions leading to the vanishing of the tadpoles in E.~\ref{eq:Vtop0}, it is useful to parametrise the phases of the 4 Yukawa couplings as follows:
\beq \label{eq:B5}
 & y_{t1} = |y_{t1} | e^{- i (\gamma_0 + \beta_0  + \varphi_0 + \delta_0)}\,, \quad 
y_{t2} = |y_{t2} | e^{- i (\gamma_0 + \beta_0  - \varphi_0 - \delta_0)}\,, & \nonumber \\
&y_{t3} = |y_{t3} | e^{- i (\gamma_0 - \beta_0  + \varphi_0 - \delta_0)}\,, \quad 
y_{t4} = |y_{t4} | e^{- i (\gamma_0 - \beta_0  - \varphi_0 + \delta_0)}\,; &
\eeq
where $\varphi_0$ and $\delta_0$ are physical phases, $\beta_0$ can be rotated away with an SU(4) rotation in Eq.~\ref{eq:Omegabeta} and $\gamma_0$ is the overall unphysical phase.
The vanishing of the tadpole for $h_2$, $Y_D^\ast Y_t = Y_D Y_t^\ast$, can be always guaranteed by a proper choice of the unphysical phase $\beta_0$:
\beq \label{equation:B9}
\beta_0 = - \delta_0 + \frac{1}{2} \mbox{arg} \left[ (|y_{t1}| - |y_{t2}| e^{2i(\delta_0 + \varphi_0)}) (|y_{t3}| - |y_{t4}| e^{2 i (\delta_0 - \varphi_0 )})\right] \;\mbox{mod} \; \pi/2\,.
\eeq
This analysis proves that the parameter $\beta$ in the vacuum is never physical, and can always be reabsorbed by a phase redefinition of the techni-fermions.

The vanishing of the tadpoles for $A_0$ and for the triplets, on the other hand, requires physical restrictions on the Yukawa couplings, which can be written in the form:
\beq \label{equation:B7}
& \left| |y_{t1}| - |y_{t2}| e^{2 i (\varphi_0 + \delta_0)} \right|  =  \left| |y_{t3}| - |y_{t4}| e^{2 i (\varphi_0 - \delta_0)} \right| \,, & \nonumber \\
& \delta_0 = - \beta_0 + \frac{1}{2} \mbox{arg} \left[ (|y_{t1}| + |y_{t4}| e^{2i(\beta_0 + \varphi_0)}) (|y_{t2}| + |y_{t3}| e^{2 i (\beta_0 - \varphi_0 )})\right] \;\mbox{mod} \; \pi/2\,, &
\eeq
where $\beta_0$ has already been fixed to cancel the $h_2$ tadpole.

The same procedure can be applied when other Yukawa couplings are added, like the bottom one.

\subsection{Non-custodial invariant vacua}

The most general vacuum can be build by rotating the EW preserving vacuum along all directions that preserve the electromagnetic U(1), and are non trivial in the EW space: this is equivalent to giving a vacuum expectation value to the two doublets and the two triplets:
\beq
\langle H_1 \rangle = \frac{1}{\sqrt{2}} \left( \begin{array}{c} 0 \\ v_1 \end{array} \right)\,, \quad  \langle H_2 \rangle = \frac{1}{\sqrt{2}} \left( \begin{array}{c} 0 \\ v_2 + i v_3 \end{array} \right)\,, \quad \langle \sigma_i \Delta^i \rangle = \langle \sigma_i N^i \rangle = \frac{v_t}{\sqrt{2}} \sigma^3\,.
\eeq
We assume that both triplets gets the same VEV, as they belong to a sextet of the custodial SO(4). A misalignment along the singlet $s$ is not interesting, as it does not touch the gauge interactions nor the Yukawa couplings~\footnote{In fact, it corresponds to a phase redefinition of the techni-fermion fields that changes the phase of the masses.}.
The most general vacuum can therefore be written as
\beq
\Sigma_{\rm gen} = e^{A}\,, \quad \mbox{with} \quad  A = \frac{1}{2\sqrt{2} f} \left( \begin{array}{cc}
i v_t\ \sigma_3 & (v_1 + i v_2)\ \matrixone + v_3\ \sigma_3 \\
- (v_1 - i v_2)\ \matrixone - v_3\ \sigma_3 & i v_t\ \sigma_3
\end{array} \right)\,.
\eeq

The vacuum can be expressed in terms of the following angles:
\beq
\tau = \frac{v_t}{2\sqrt{2} f}\,, \quad \tan \beta_1 = \frac{v_2}{v_1+v_3}\,,\quad \tan \beta_2 = \frac{v_2}{v_1 - v_3}\,,
\eeq
and
\beq
\theta_+ = \frac{\sqrt{(v_1 + v_3)^2 + v_2^2}}{2\sqrt{2} f}\,, \quad \theta_- = \frac{\sqrt{(v_1 - v_3)^2 + v_2^2}}{2\sqrt{2} f}\,, \quad \mbox{with}\;\; \frac{\theta_+^2}{\theta_-^2} = \frac{1+\tan^{-2} \beta_1}{1+\tan^{-2} \beta_2}\,.
\eeq
The result reads
\beq \label{eq:general vacuum}
\Sigma_{\rm gen} = \left( \begin{array}{cccc}
e^{i \tau} \cos \theta_+ & 0 & e^{i (\beta_1 + \tau)} \sin \theta_+ & 0 \\
0 & e^{-i \tau} \cos \theta_- & 0 & e^{i (\beta_2 - \tau)} \sin \theta_- \\
- e^{i (-\beta_1 + \tau)} \sin \theta_+ & 0 & e^{i \tau} \cos \theta_+  & 0 \\
0 & e^{i (-\beta_2 - \tau)} \sin \theta_- & 0  & e^{-i \tau} \cos \theta_- 
\end{array} \right)\,.
\eeq
The limit of custodial vacuum can be reached by setting $\tau = 0$, $\theta_+ = \theta_- = \theta$, and $\beta_1 = \beta_2 = \beta$~\footnote{Note that for $v_2 = 0$, the two phases $\beta_1$ and $\beta_2$ vanish, however $\theta_+/\theta_-$ is not determined and they should be considered as independent free parameters.}. It can also be shown that the above vacuum is equivalent to one constructed starting from the custodial invariant one, to which a rotation along the imaginary part of the second doublet and along the neutral triplets is applied.

The masses of the $W$ and $Z$ are now given by:
\beq \label{eq:boson mass}
m_W^2 &=& 2 g^2 f^2 (1-\cos (2\tau)\ \cos \theta_+\ \cos \theta_- )\,, \\
m_Z^2 &=& (g^2 + {g'}^2) f^2 (\sin^2 \theta_+ + \sin^2 \theta_-)\,,
\eeq 
while the Weinberg angle has the same value as in the SM, $\cos^2 \theta_W = g^2/(g^2+{g'}^2)$.
The tree level correction to the $\rho$ parameter is thus given by
\beq
\rho - 1 = \frac{m_W^2}{m_Z^2 \cos^2 \theta_W} - 1 = \frac{(\cos \theta_+ - \cos \theta_-)^2 + 4 \sin^2 \tau\ \cos \theta_+\ \cos \theta_-}{\sin^2 \theta_+ + \sin^2 \theta_-}\,.
\eeq
To have a quantitative idea of the constraints coming from $\rho$, which is close to 1 up to $\sim 10^{-3}$, it is useful to expand for small breaking of the custodial symmetry:
\beq
\rho - 1 \sim \frac{(\theta_+ - \theta_-)^2}{2} + 2 \frac{\tau^2}{\tan^2 \theta}\,,
\eeq
thus the constraint on $\theta_+ - \theta_-$ is of the order of few times $10^{-2}$, while the contribution of the triplet is enhanced by a small $\theta$ and is thus stronger.

We notice that, in the vacuum in Eq.~\ref{eq:general vacuum}, the triplets and the real component of the second doublet enter like phases, respectively $\tau$ and the $\beta_{1,2}$ pair. It is therefore instructive to investigate the relation with the Yukawa phases. 
We can re-write the general vacuum as
\begin{eqnarray}
 \Sigma_{\rm gen} = U_\tau\cdot \Sigma_{\rm gen} ^\prime \cdot U_\tau   \nonumber \\
\end{eqnarray}
with 
\beq
\Sigma_{\rm gen} ^\prime =  \left( \begin{array}{cccc}
 \cos \theta_+ & 0 & e^{i \beta_1 } \sin \theta_+ & 0 \\
0 & \cos \theta_- & 0 & e^{i \beta_2 } \sin \theta_- \\
- e^{-i \beta_1 } \sin \theta_+ & 0 & \cos \theta_+  & 0 \\
0 & -e^{-i \beta_2 } \sin \theta_- & 0  & \cos \theta_- 
\end{array} \right)\,.
\eeq
The vacuum $\Sigma'_{\rm gen}$ contains the contribution of the two doublets, while the matrix $U_\tau$, generated by the triplets, is given by 
\begin{eqnarray}
U_\tau= \left( \begin{array}{cc}
e^{i\frac{\tau \cdot\sigma_3}{2} }   & 0   \\
0 & e^{i\frac{\tau \cdot\sigma_3}{2} } \\
\end{array} \right)\,.
\end{eqnarray}   
From the general top Yukawa coupling in Eq.~\ref{eq:Yuk0}, it is the coupling of the tops that determines the tadpole for the triplets in the loop induced potential: such coupling can be written as
\begin{eqnarray} 
-f(\bar{\chi}_{t_L} \bar{\eta}_{t_R}) \{ Tr[P_{1,1} (y_{1t}e^{i\tau }\Sigma_{gen} ^\prime +y_{2t}e^{-i\tau }  \Sigma_{gen} ^{\prime \dagger})] +i\sigma_2  ^{1, 2}Tr[P_{2} ^2 ( y_{3t}e^{-i\tau }\Sigma_{gen}  ^\prime  + y_{4t}e^{i\tau } \Sigma_{gen} ^{\prime \dagger}) ]     \}  +h.c \nonumber \\ 
\label{equation:B4}
\end{eqnarray}
where we have replaced the pNGB matrix with the new vacuum $\Sigma_{\rm gen}$, and explicitly written down the contribution of the phase induced by the triplets.
Then, we can re-write  the Yukawa couplings as   
\beq 
 & y_{t1} ^\prime =y_{t1}e^{i\tau }= |y_{t1} | e^{- i (\gamma_0 + \beta_0  + \varphi_0 + (\delta_0 -\tau))}\,, \quad 
y_{t2}  ^\prime =y_{t2}e^{-i\tau}= |y_{t2} | e^{- i (\gamma_0 + \beta_0  - \varphi_0 - (\delta_0 -\tau))}\,, & \nonumber \\
&y_{t3}  ^\prime=y_{t3}e^{-i\tau } = |y_{t3} | e^{- i (\gamma_0 - \beta_0  + \varphi_0 - (\delta_0 -\tau))}\,, \quad 
y_{t4}  ^\prime=y_{t4}e^{i\tau } = |y_{t4} | e^{- i (\gamma_0 - \beta_0  - \varphi_0 + (\delta_0-\tau))}\,. &
\eeq
The above equations show therefore that the vacuum along the triplet direction corresponds to a redefinition of the phase $\delta_0$ in the 4 Yukawa couplings: as such a phase is physical (i.e., it cannot be removed by a phase redefinition of the techni-fermion fields), it is $\delta_0 - \tau$ that contains the physical effect of the triplet vacuum.
This analysis also confirms that the tadpole for the triplet can be eliminated by appropriately fixing the value of $\delta_0$, as in Eq.~\ref{equation:B7}.

The two phases generated by the second doublet vacuum,  $\beta_1 $ and $\beta_2$, can be rotated away by two rotations $\Omega_\beta$ in Eq.~\ref{eq:Omegabeta} and  $\Theta_{ \varphi   }$:
\begin{eqnarray}
(\Theta_{ \varphi   }\cdot \Omega_\beta)\cdot \Sigma_{\rm gen}\cdot ( \Theta_{\varphi} \cdot \Omega_\beta )^\dagger  =U_\tau\cdot\left( \begin{array}{cccc}
 \cos \theta_+ & 0 & \sin \theta_+ & 0 \\
0 &  \cos \theta_- & 0 &\sin \theta_- \\
 -\sin \theta_+ & 0 &  \cos \theta_+  & 0 \\
0 & -\sin \theta_- & 0  &  \cos \theta_- 
\end{array} \right)\cdot U_\tau 
\end{eqnarray}  
with $\varphi =\frac{\beta_1 -\beta_2}{2} $, $\beta = \frac{\beta_1 +\beta_2}{2} $. $ \Theta_{ \varphi   } $ corresponds to a local phase transformation generated by the generator of the Hypercharge:
\begin{eqnarray}
 \Theta_{ \varphi   } =\left( \begin{array}{cc}
\bf{ 1_{2\times 2}}   & \bf{0_{2\times 2}} \\
\bf{0_{2\times 2}}& \left( \begin{array}{cc}
e^{i\varphi} & 0 \\
0 & e^{-i\varphi} \\
 \end{array} \right)  \\
\end{array} \right)\,.
\end{eqnarray} 
So the kinetic terms are independent on the phases $\beta_1 $  and $ \beta_2$, and thus the mass of  $W^{\pm} _\mu $ and $Z_\mu $ in Eq.~\ref{eq:boson mass}. As $\Theta_{ \varphi } $ is only a global hypercharge  $U(1)_Y$ transformation, $\varphi=\frac{ \beta_1-\beta_2}{2} $ is an unphysical [hase. We already know  that  $\Omega_\beta $ is simply a redefinition of techni-fermion unphysical phase, which can transfer the phase $\beta=\frac{\beta_1 +\beta_2}{2} $ from the vacuum  to the fermion Yukawa couplings.  So  we can always use this freedom  to set  $\beta=\frac{\beta_1 +\beta_2}{2}  $ in vacuum  to be zero which  is  equivalent to set  both  $\beta_{1,2} $ be zero  and also means we can always  set  $\upsilon_2 =0 $.

\section{Mass matrices for the pions}
\label{app:masses}
\setcounter{equation}{0}
\setcounter{footnote}{0}

The pNGBs in the theory receive mass contributions from the 3 terms in the potential:
\beq
M_{\varphi}^2 = \Delta M^2_{M} + \Delta M^2_{G} + \Delta M^2_{T}\,,
\eeq
respectively coming from the techni-fermion mass, the gauge loops and top loops.

The contribution of the TQ mass--induced potential, Eq.~\ref{eq:Vmass}, gives diagonal and gauge invariant masses to the pNGB multiplets
\beq
\Delta M^2_{h_1} = \Delta M^2_{H_2} = \Delta M^2_s &=& \frac{C_m (m_L+m_R) f}{2} c_\theta\,, \\
\Delta M^2_{\Delta}  &=& \frac{C_m (m_L+m_R) f}{2}  (c_\theta + \delta)\,, \\
\Delta M^2_{N}  &=& \frac{C_m (m_L+m_R) f}{2}  (c_\theta - \delta)\,;
\eeq
where the subscript $H_2$, $\Delta$ and $N$ indicate the common mass of the second doublet, and the two triplets respectively, and we have defined
\beq
\delta = \frac{m_L - m_R}{m_L + m_R}\,.
\eeq

Loops of gauge bosons in Eq.~\ref{eq:Vgauge}, on the other hand, give different mass contribution to the components of the multiplets, as the vacuum is not gauge invariant. Notably, the masses are diagonal in the gauge-basis, except for a mixing between the two triplets:
\beq
\Delta M^2_{h_1} &=& \frac{C_g f^2}{8} (3 g^2 + {g'}^2) c_{2\theta}\,, \\
\Delta M^2_{h_2} &=& \frac{C_g f^2}{8} (3 g^2 + {g'}^2) c^2_{\theta}\,, \\
\Delta M^2_{A_0} &=& \Delta M^2_{h_2} + \frac{C_g f^2}{8} \left(g^2 - {g'}^2\right) s_\theta^2\,, \\
\Delta M^2_{H^\pm} &=& \Delta M^2_{h_2} + \frac{C_g f^2}{8} \left(g^2  + {g'}^2 \right) s_\theta^2\,, \\
\Delta M^2_{s} &=& - \frac{C_g f^2}{8} (3 g^2 + {g'}^2) s^2_{\theta}\,;
\eeq
and the triplet masses, expressed as 2$\times$2 matrices in the $\Delta$--$N$ basis:
\beq
\Delta M^2_{\Delta_0-N_0} &=&  \frac{C_g f^2}{8} \left\{ 8 g^2 \left( \begin{array}{cc}
 c_{\theta/2}^2 & 0 \\ 0 & s_{\theta/2}^2 \end{array} \right) -\frac{1}{2} \left( \begin{array}{cc}
7 g^2 + {g'}^2 & g^2 - {g'}^2 \\ g^2 - {g'}^2  & 7 g^2 + {g'}^2 \end{array} \right) s_\theta^2 \right\}\,, \\
\Delta M^2_{\Delta^\pm-N^\pm} &=& \Delta M^2_{\Delta_0-N_0} + \frac{C_g f^2}{8}  {g'}^2 \left( \begin{array}{cc}
(1-c_\theta)^2 & s_\theta^2 \\ s_\theta^2 & (1+c_\theta)^2 \end{array} \right)\,.
\eeq
The above formulas explicitly show that in the limit $\theta \to 0$, the multiplets receive a common mass, while in the triplet $N$ only the charged components receives a correction from the hypercharge gauge boson.

The contribution of top loops to the pNGB masses (and potential) is in Eq.~\ref{eq:Vtop}. After imposing the conditions on the pre-Yukawas that ensure a custodial invariant vacuum, i.e. $Re (Y_D) = 0$ and $Im (Y_T) = 0$, and using the arbitrary phase redefinitions to fix $Y_t$ real and $\beta = 0$ (i.e., $Im (Y_D) = 0$), the mass matrices depend on 4 independent parameters: the top Yukawa $Y_t$, $Y_T$ and the complex parameter $Y_0$.
Contrary to the other contributions, the top loops generate mixing among all the neutral pNGBs (expect the Higgs-like state $h_1$), and the charged ones.
The mass correction to $h_1$ is given by:
\beq
\Delta M^2_{h_1} = - 2 C_t f^2 Y_t^2 c_{2\theta}\,.
\eeq
The singlet $s$ is also left unmixed:
\beq
\Delta M^2_{s} = 2 C_t f^2 Y_t^2 s_\theta^2\,.
\eeq
The contribution to the neutral masses, in the basis $\{A_0, h_2, \Delta_0, N_0\}$, and writing $Y_0 = Y^R_0 + i Y^I_0$ is given by:
\beq
\Delta M^2_{0} = 2 C_t f^2 \left( \begin{array}{cccc}
Y_t^2\ s_\theta^2 & 0 & \frac{1}{2 \sqrt{2}} Y_t Y^I_0 \ s_{2\theta} &  \frac{1}{2 \sqrt{2}} Y_t Y^I_0\ s_{2\theta} \\
 0 &  Y_t^2\ s_\theta^2 & \frac{1}{\sqrt{2}} Y_t Y^R_0\ s_\theta &  \frac{1}{\sqrt{2}} Y_t Y^R_0\ s_\theta   \\
\frac{1}{2 \sqrt{2}} Y_t Y^I_0\ s_{2\theta}    &  \frac{1}{\sqrt{2}} Y_t Y^R_0\ s_\theta &  \left(Y_t^2 - \frac{1}{2} Y_T^2 \right)\ s_\theta^2 & - \frac{1}{2} Y_T^2\ s_\theta^2 \\
 \frac{1}{2 \sqrt{2}} Y_t Y^I_0\ s_{2\theta}  &  \frac{1}{\sqrt{2}} Y_t Y^R_0\ s_\theta & -\frac{1}{2} Y_T^2\ s_\theta^2 &   \left(Y_t^2 - \frac{1}{2} Y_T^2 \right)\ s_\theta^2 
 \end{array} \right)\,.
\eeq
It is interesting to notice that in the case of real pre-Yukawas, $Y_0^I = 0$, $A_0$ decouples from the other neutral scalars, due to the different CP properties of the fields: in fact, $A_0$ is CP-even in this limit.
On the other hand, for a purely imaginary $Y_0$, i.e. $Y_0^R = 0$, it is $h_2$ that decouples: in this limit therefore, one can redefine the CP properties of the pNGBs so that $h_2$ is CP even and $A_0$ CP-odd.
In the charged sector, in the basis $\{H^\pm, \Delta^\pm, N^\pm\}$, we have
\beq
\Delta M^2_{\pm} = 2 C_t f^2 \left( \begin{array}{ccc}
  Y_t^2\ s_\theta^2 & -\frac{1}{\sqrt{2}} Y_t (Y^R_0 + i Y_0^I c_\theta)\ s_\theta & Y_t^2 \frac{1}{\sqrt{2}} Y_t (Y^R_0 - i Y_0^I c_\theta)\ s_\theta \\
-\frac{1}{\sqrt{2}} Y_t (Y^R_0 - i Y_0^I c_\theta)\ s_\theta & \left( Y_t^2 - \frac{1}{2} Y_T^2 \right)\ s_\theta^2 & - \frac{1}{2} Y_T^2\ s_\theta^2  \\
\frac{1}{\sqrt{2}} Y_t (Y^R_0 + i Y_0^I c_\theta)\ s_\theta & - \frac{1}{2} Y_T^2\ s_\theta^2  &  \left( Y_t^2 - \frac{1}{2} Y_T^2 \right)\ s_\theta^2  
\end{array} \right)\,.
\eeq

\end{document}